\documentclass[conference]{IEEEtran}
\IEEEoverridecommandlockouts
\usepackage{cite}
\usepackage{amsmath,amssymb,amsfonts}
\usepackage{algorithmic}
\usepackage{graphicx}
\usepackage{textcomp}
\usepackage{xcolor}
\usepackage{hyperref}
\usepackage{multirow}
\usepackage{longtable}
\usepackage{array}
\def\BibTeX{{\rm B\kern-.05em{\sc i\kern-.025em b}\kern-.08em
    T\kern-.1667em\lower.7ex\hbox{E}\kern-.125emX}}
\usepackage{comment}
\usepackage[font=small,labelfont=small]{caption}
\usepackage{subcaption}
\usepackage{soul}

\newcommand\teq{\mkern1.5mu{=}\mkern1.5mu}
\newcommand{\nonsummary}[6]{{\small $p\teq{#1}, g^*\teq{#4}, CI_{.95}[{#5},{#6}]$}}
\newcommand{\summary}[5]{{\small$g^*\teq{#3}, CI_{.95}[{#4},{#5}]$}}

\newcommand{\figref}[1]{Fig.~\ref{#1}}
\newcommand{\trans}[2]{{#1}$\rightarrow${#2}}

\newcolumntype{L}[1]{>{\raggedright\let\newline\\\arraybackslash\hspace{0pt}}p{#1}}
\newcolumntype{C}[1]{>{\centering\let\newline\\\arraybackslash\hspace{0pt}}p{#1}}

\begin{document}

\title{Error-related Potential Variability: Exploring the Effects on Classification and Transferability}

\author{\IEEEauthorblockN{ Benjamin Poole\IEEEauthorrefmark{1} and Minwoo Lee\IEEEauthorrefmark{2}}  
\IEEEauthorblockA{\textit{The Department of Computer Science} \\
\textit{University of North Carolina at Charlotte}\\
Charlotte, NC, USA \\
bpoole16@uncc.edu\IEEEauthorrefmark{1}, minwoo.lee@uncc.edu\IEEEauthorrefmark{2}}
}

\maketitle

\begin{abstract}
Brain-Computer Interfaces (BCI) have allowed for direct communication from the brain to external applications for the automatic detection of cognitive processes such as error recognition. Error-related potentials (ErrPs) are a particular brain signal elicited when one commits or observes an erroneous event.  However, due to the noisy properties of the brain and recording devices, ErrPs vary from instance to instance as they are combined with an assortment of other brain signals, biological noise, and external noise, making the classification of ErrPs a non-trivial problem. Recent works have revealed particular cognitive processes such as awareness, embodiment, and predictability that contribute to ErrP variations. In this paper, we explore the performance of classifier transferability when trained on different ErrP variation datasets generated by varying the levels of awareness and embodiment for a given task. In particular, we look at transference between observational and interactive ErrP categories when elicited by similar and differing tasks. Our empirical results provide an exploratory analysis into the ErrP transferability problem from a data perspective.
\end{abstract}

\section{Introduction}

The rise of brain-computer interfaces (BCIs) has brought about a direct line of communication from the brain to external devices. This has largely been thanks to non-invasive devices like electroencephalographs (EEGs) which have increased the accessibility of BCIs, thus opening up a wider variety of applications. As EEGs have a high temporal resolution at the level of milliseconds\cite{luck_erpbook_2014}, they allow for the observation of the brain's response to events and stimuli which can reflect certain cognitive processes such as error recognition \cite{falkenstein_ern_2000, hoffmann_errpage_2012}. EEGs are particularly useful at capturing a type of brain signals referred to as \textit{event-related potentials} (ERPs) which are elicited in response to various internal or external stimuli and characterized by their unique temporal signal structure of peaks and troughs. A subset of ERPs elicited when one perceives or commits erroneous action are referred to as error-related potentials (ErrPs). ErrPs are generally characterized by a negative peak, often referred to as error-related negativity (ERN), that occurs between 80-300~ms and a subsequent positive peak (Pe) that occurs between 200-500~ms \cite{falkenstein_ern_2000, luck_erpbook_2014, holroyd_frn_2002, ferrez_errp_2008, spuler_continuouserrp_2015}. Recently, ErrPs have been of particular interest for BCI applications, ranging from simple detection/classification of errors \cite{ferrez_errpwrong_2005,ferrez_errp_2008, spuler_continuouserrp_2015}, to training robots and agents \cite{xu_accelerating_2020, wang_maximizing_2020}, to correcting mistakes made by BCI-controlled interfaces \cite{chavarriage_errpbci_2014}. 

However, inherent to the classification of ErrPs is the problem of constant variations in the observed signal structure.
Collected neural activity data is quite noisy as it consists of the accumulation of processes occurring both internally and externally. For instance, any given brain signal is accompanied by a great deal of biological noise introduced through cognitive processes occurring within the mind  (e.g., fatigue, attention, engagement) and biological artifacts (e.g., eye blinks, muscle movements). At the time of signal acquisition, external noise is additionally introduced through external electrical activity (e.g., power lines, electronics). Thus, detecting ErrPs is difficult as the captured neural activity not only contains the ErrPs signal but a host of ever-changing unrelated signals. 

Nevertheless, common ErrP variations (i.e., augmentations in the observed ErrP signal structure) have been able to be detected, likely due to the underlying invariant features for these ErrPs \cite{abu_errpinvariance_2017}. For instance, \textit{observation} are elicited in response to observed erroneous action\cite{ferrez_errp_2008}; \textit{feedback} are elicited when feedback indicates an erroneous choice or action was made\cite{holroyd_frn_2002}; \textit{outcome} are elicited when the goal of an action is not achieved\cite{milekovic_error-related_2012, spuler_continuouserrp_2015}; \textit{execution} or \textit{interaction}\footnote{Execution ErrPs are elicited with continuous actions while interaction ErrPs are elicited with discrete actions \cite{spuler_continuouserrp_2015}} are elicited when an action is input to an interface and the interface outputs an erroneous action instead \cite{milekovic_error-related_2012, spuler_continuouserrp_2015, ferrez_errp_2008}.

A variety of causes are thought to be connected to these variations, such as the influence of cognitive processes elicited due to a particular task \cite{abu_errpinvariance_2017, ehrlich_errpfeasibility_2019, iturrate_errptask_2013}, changes in task complexity \cite{kakkos_errpcomplex_2020}, repeated elicitation of the ErrP signal \cite{ferrez_errp_2008}, and changes in the way a task is interacted with \cite{kim_handlingtransfer_2016}. In more general terms, the notion of awareness, embodiment, and predictability seem to account for some of the noticeable differences in ErrP variations. Here, \textit{awareness} refers to the conscious perception of errors such that an increase in conscious awareness of a particular error leads to larger Pe peaks \cite{iwane_invariability_2021, orr_role_2011}. \textit{Embodiment} relates to whether a subject interacts with the task in 1st or 3rd person where lower levels of embodiment (3rd person observation) lead to weaker ERN amplitudes and higher levels of embodiment (1st person interaction) lead to stronger ERN amplitudes \cite{pavone_embodying_2016}. \textit{Predictability} is thought to augment the latency of ErrP components where errors that are considered more ``unpredictable" increase latency \cite{iwane_invariability_2021}. 

 The constant variation in observed brain signals leads to the difficult problem, often seen in machine learning, of non-stationarity. To further reduce this problem, BCI applications usually entail tedious calibration periods where subject-specific classifiers are created \cite{huang_review_2021}. In order to create these subject-specific classifiers, ideally, before any real experiment begins, the target task is used to generate data for the initial training of the classifier. Unfortunately, the target tasks are often expensive to collect data from, thus requiring a simplified source task to provide data for calibration.
 
 One approach to handling calibration when the source and target task differ is to investigate the algorithmic side. This becomes a machine learning focused question where we ask how can we build better algorithms that can properly transfer knowledge (i.e., transfer learning) \cite{azab_bcitransfer_2018}. Likewise, we can also ask a similar question specific to the task, which task might produce the ``optimal" data that can train an initial classifier to work effectively with the target task data? It is then of interest to understand what source-target task design might be needed such that there is a minimal drop in classifier performance. Additionally, it would be desirable to have a controlled and simplistic source task that allows large quantities of data to be gathered quickly.
\vspace{-.6cm}
\subsection{Research Questions}

The objective of this paper is to explore the transferability of two general categories of ErrP variations. In particular, we are interested in what we define as observational and interactive ErrP categories. We use the \textit{observational} category to refer to any ErrP variation elicited while observing an erroneous event. This definition corresponds closely with the observation ErrP defined in prior works \cite{ferrez_errp_2008}. Based on prior studies, we hypothesize that observational ErrPs are ``less" distinct\footnote{Distinct here refers to the increase or decrease in amplitudes in the ErrP signal structure such as the ERN and Pe.} such that they have lower levels of embodiment and, potentially, awareness since they entail having a subject simply observe a task \cite{pavone_embodying_2016}. Meanwhile, we use the \textit{interactive} category to refer to any ErrP variation elicited while interacting with a task. This category entails multiple types of ErrP variations such as interaction, execution, and outcome. We hypothesize that interactive ErrPs are ``more'' distinct due to higher levels of embodiment and, potentially, awareness from subjects directly interacting and engaging with the task. We establish the following research questions to guide our exploration of ErrP transferability, and in turn, provide insight into our prior hypotheses as well.
\begin{enumerate}
  \item \textbf{Q1}: Do more ``distinct'' ErrP variations produce ``easier'' classification problems? 
  \item \textbf{Q2}: What role does the category of ErrP variation or task similarity play in transfer performance?
  \item \textbf{Q3}: Following Q1 and Q2, what practical implications can be extracted? 
\end{enumerate}

Unlike prior studies which typically focus on specific ErrP variations within the same task, this work aims to look at the classification of ErrP categories within and between different tasks as this is often a requirement for calibration. 

\section{Experiment Design for Analysis}\label{sec:experiments}


\subsection{Tasks}

Simple discrete and controlled tasks enable a clear distinction between correct and erroneous events \cite{abu_errpinvariance_2017, iturrate_errptask_2013, ferrez_errp_2008, ferrez_errpwrong_2005}. This allows for a straightforward assessment of ErrP variations from the perspective of classification and visualization of grand average waveforms. In this paper, we reuse this idea of a simple and controlled task but also investigate slightly more uncontrolled, complex, and continuous tasks where ErrPs are less clearly defined (i.e., lower signal-to-noise ratio and noisier ground truth), closer to our real-world scenarios.  We design two tasks, Binary Goal Search (BGS) and Obstacle Avoidance (OA), to test the potential of within/between task and ErrP category transfer. Each task is broken into a sub-task where each sub-task corresponds to our previously defined categories of ErrP variations. Thus, BGS and OA tasks both have two sub-tasks: observational and interactive. 

\begin{figure}[t] 
    \begin{subfigure}[t]{.5\columnwidth}
        \centering
        \raisebox{0.2\height}{\includegraphics[width=\linewidth]{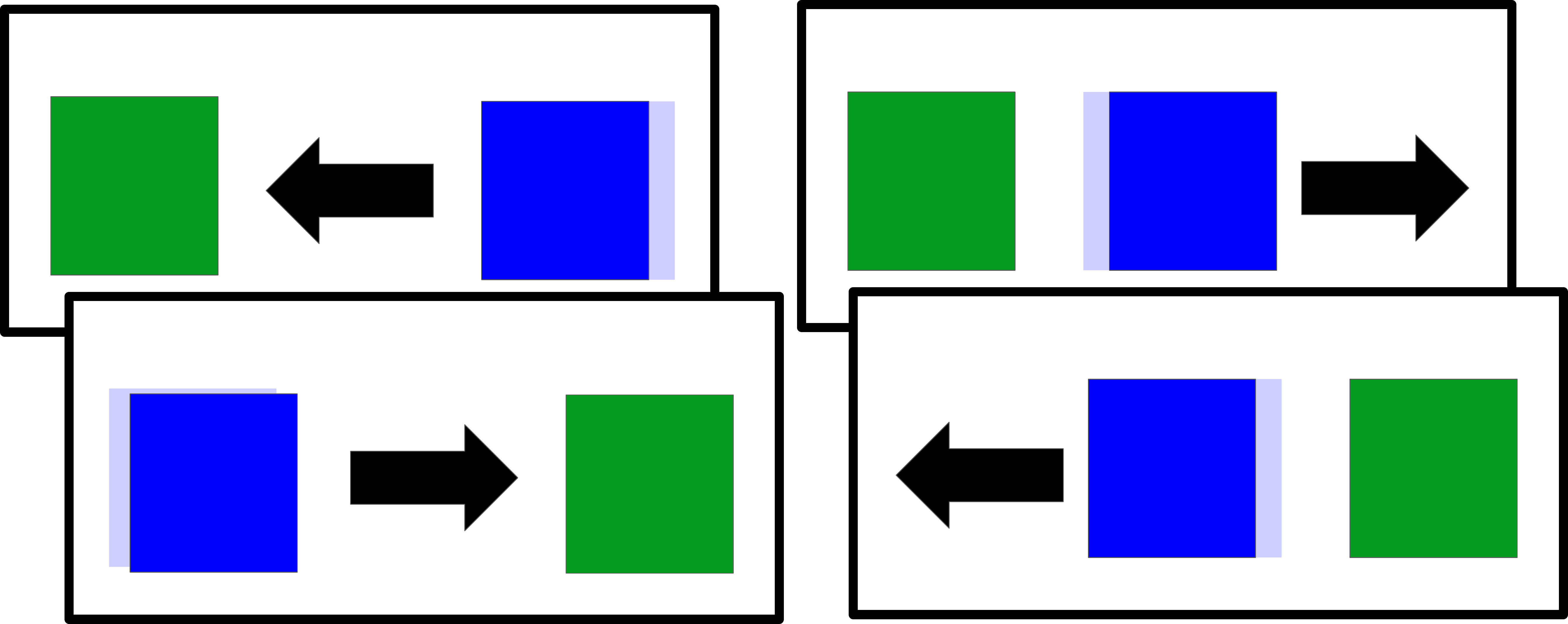}}
        \caption{}
        \label{fig:bgs_task}
    \end{subfigure}
    \hfill
    \begin{subfigure}[t]{.4\columnwidth}
        \centering
        \framebox{\includegraphics[width=.75\linewidth]{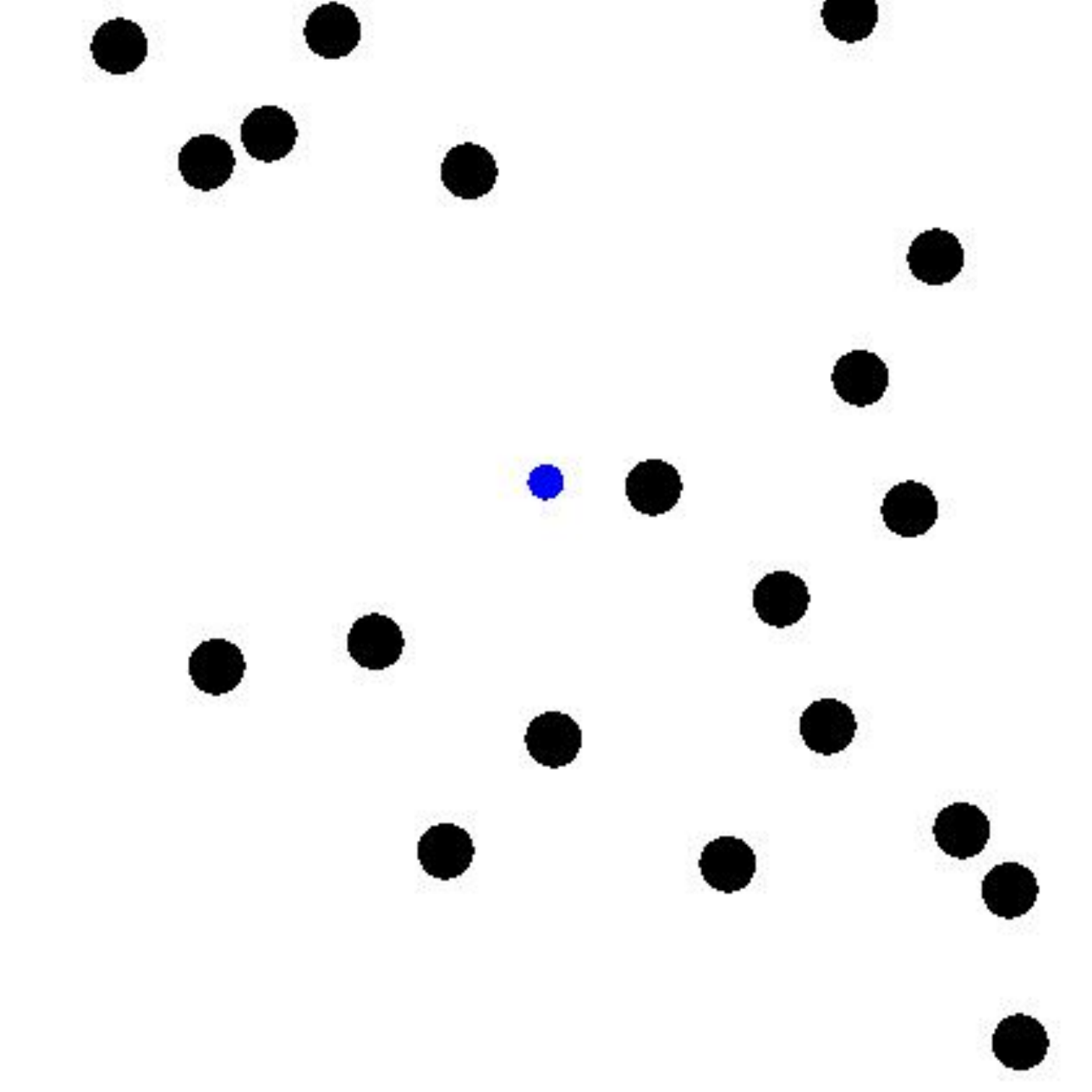}}
        \caption{}
        \label{fig:oa_task}
    \end{subfigure}
    \caption{Experiment tasks where (\subref{fig:bgs_task}) shows BGS task (the right column shows the correct actions and the left column shows incorrect actions) while (\subref{fig:oa_task}) shows the OA task.}
    \vspace{-0.5cm}
\end{figure}

\subsubsection{BGS Task}

The objective of the BGS task is for the player (i.e., human/agent) to move towards a goal position generated on the screen (\figref{fig:bgs_task})\footnote{A GIF of the BGS task can be viewed at \url{https://tinyurl.com/2whth2jd}}. The blue square is controlled by the player, while the green square represents the goal. At the start, the goal spawns on either the right or left hand side of the blue square. The blue square spawns in the middle of the screen, three steps away from the goal. To properly discretize this task (i.e. keep ErrPs overlapping from occurring), there is approximately 1.2 s $\pm 100$ ms delay between each action as recommended in \cite{luck_erpbook_2014}. As the player moves towards the goal, there is a 20\% chance for an error to occur where the interface moves the player in wrong direction. When this occurs, an ErrP is expected to be elicited. A 20\% error rate is chosen inline with prior works \cite{iturrate_errptask_2013, ferrez_errpwrong_2005, ferrez_errp_2008}. Once the goal has been reached, the player's blue square is reset to the center of the screen and a new goal randomly spawns. Finally, if the player is ever on the edge of the screen, an error will never occur and they will always move in the correct direction as this prevents the player from moving off the screen.

For the observational sub-task (BGSObs), a predefined agent is used to control the blue square. The agent will always move towards the goal with a 20\% chance of error. When an error occurs, an observation ErrP is elicited. Likewise, for the interactive sub-task (BGSInt), the human controls the blue square. Every time the task is ready to receive an input the human must enter either ``a" for moving right or ``d" for moving left. Once again, there is 20\% chance of error\footnote{Errors can only occur if the human selects the correct action (i.e., towards the goal) to begin with.} occurring in which the interface will move the human in the wrong direction away from the goal. When an error occurs, an interaction ErrP is elicited. Trials in which the human chooses the wrong direction intentionally or on accident are still marked as errors as they correspond to the incorrect action.

The BGS task is used to represent a controlled environment. Variations of the BGS task have frequently been used to elicit ErrPs by prior works \cite{ferrez_errpwrong_2005, ferrez_errp_2008, abu_errpinvariance_2017}. The BGS tasks are desirable as we can regulate the expected number of ErrPs elicited. Furthermore, a simple environment containing only binary actions presents a clear right or wrong choice. Such traits are desirable to quickly collect data to pre-train a classifier. For instance, if the target task is more complex and error elicitation is sparse, a simple task could be more desirable to collect data quickly while controlling error elicitation. 

\subsubsection{OA Task}

Inspired by the popular mobile game Flappy Bird, the objective of this task is for the player to avoid obstacles that move horizontally across the screen (\figref{fig:oa_task})\footnote{A GIF of the OA task can be viewed at \url{https://tinyurl.com/4td2uj3d}}. The player controls the blue circle that spawns in the center of the screen and must avoid colliding with obstacles (black circles) and the boundaries of the window. The player must avoid obstacles by moving up (i.e., increase y-coordinate) or letting gravity drag them down (i.e., decreasing y-coordinate). The player moves up by clicking the left mouse button. The player moves down by performing no action as a crude form of simulated gravity automatically pulls them down towards the bottom of the screen. Obstacles spawn on the right-hand side of the screen and move to the left-hand side. If the player collides with an obstacle, the game will freeze for 1 second and the player's blue circle turns red. This 1 second delay is used to ensure the player recognizes they have collided with an object and to allow a sufficient duration for the ErrP to be captured in isolation. Once the game resumes the player remains red for 1 more second and is considered immune such that colliding with an obstacle does nothing. Subjects are informed of this mechanic ahead of time but not told if collision during this period is explicitly correct or erroneous, thus this is a gray area where ErrPs might be elicited depending on the subject's subjective take on the immune state\footnote{As such, labels for collisions are not reported during the immune state.}.

For the observational sub-task (OAObs), a deep Q-network agent \cite{mnih_dqn_2015} is trained to control the player object  where trajectories from the agent are shown to subjects. We manually select a sub-optimally trained agent that can successfully perform the task but which also commits frequent errors, typically more than the human subjects. When an known error occurs, an observation ErrP is elicited. Likewise, the interactive sub-task (OAOut) lets the human control the player object. When a known error occurs, an outcome ErrP is elicited.

Unlike the BGS task, the error rate of the OA task is highly dependent on the subject's skill at playing the game, representing a more uncontrolled environment. Game difficulty can be increased to ensure more collisions occur by increasing the number of obstacles that spawn per second and adjusting the gravity pulling on the player object. However, this still does not ensure an expected number of errors per trial and subject. Thus, ErrP data for this task is often more sparse and tends to become increasingly more sparse as subjects play more and more trials as they get better at the game. Additionally, this task can introduce a noisy ground truth where only \textit{known} ErrPs are labeled. That is, we can only label ErrPs approximately where we think they might occur (e.g., when the player collides with an object). Thus, we cannot account for other errors a subject might perceive. Furthermore, we cannot account for when an ErrP is elicited just before the collision as the player foresaw a collision occurring \cite{spuler_continuouserrp_2015}. Finally, we suspect the OA task will produce nosier data due to an increase in cognitive and biological functions required to perform the task although these aspects are not quantified explicitly. For instance, noise from eye movements and saccades are more likely as subjects must track the player object and obstacles all over the screen instead of remaining in a relatively fixed position. We also suspect a higher degree of decision making and planning are required for this task as it is more complex than the BGS task \cite{kakkos_errpcomplex_2020}.

\subsection{Data Preparation and Classifiers}

A total of 9 healthy subjects (2 female, 7 male) participated in the data collection. Each subject had 2 recording sessions where a single session contained 5 trials of each sub-task. A single trial for all tasks took approximately 3-4 minutes to complete. Thus, a total of 40 trials for each subject were recorded. All subjects were seated approximately 7 feet away from a large 4k display that presented the tasks in the following order: BGSObs, BGSInt, OAObs, OAOut. The instructions for each sub-task were presented before the task began. For instance, subjects were instructed that moving towards the goal in the BGS task was considered correct while colliding with objects in the OA task, except during the immune period, was considered erroneous. Each observational sub-task was presented before the interactive sub-task so subjects could see the task before performing it themselves. Before each task began, a 30-second resting period was allotted. Additionally, subjects were allowed to rest for 10-15 minutes between each task if needed.

Neural activity was recorded using a Ganglion Board from OpenBCI which has a max capacity of four electrodes and a sampling rate of 200 Hz. The electrodes are held in place via a 3D printed frame provided by OpenBCI that follows the 10-20 electrode placement system. The four electrodes are placed at the F3, F4, Fz, and Cz positions as ErrPs are thought to be elicited from the anterior cingulate cortex in the medial pre-frontal region \cite{falkenstein_ern_2000}. To compensate for limited data, we utilize asynchronous epoching where we pass a sliding window over the data such that the labels are mapped to the closest window. Here the sliding window is 700 ms long, where we use a step size of 100 ms such that there is approximately 90\% overlap per window. This allows us to increase the amount of training data drastically at the cost of further violating the idea of independence in the dataset. In order to ground truth the data, we map labels to their corresponding window such that each window has the label NoErrP/ErrP.


For our choice of classification algorithms, we utilize deep neural networks (DNN) and support-vector machines (SVM) as they are two popular classification algorithms for classifying ERPs and ErrPs \cite{lotte_bcisurvey_2018, lawhern_eegnet_2018}. In particular, we use EEGNet, a deep convolutional neural network \cite{lawhern_eegnet_2018}. EEGNet is well known for being extremely compact and having few parameters while still being effective at classifying ErrPs. Thus, it allows for learning from a relatively small amount of data as the cost of collecting EEG data is often expensive. We implement EEGNet in TensorFlow and utilize the default hyper-parameters corresponding to EEGNet-8-2 \cite{lawhern_eegnet_2018}. We utilize scikit-learn's SVM implementation with a linear kernel and a regularization parameter of 1.0. 

All data is preprocessed using an IIR 0.1-30 Hz order 4 band-pass filter. The remaining data preprocess steps vary slightly depending on the classifier. For EEGNet, data is scaled using z-standardization and all data is resampled to 128 Hz in accordance to \cite{lawhern_eegnet_2018}. Finally, data is upsampled such that the number of ErrP data samples matches the number of NoErrp data samples. For the SVM, the data is only scaled using min-max scaling and no up-sampling is performed. Instead, the data imbalance is compensated for by using class weighting where the weight of the classes is determined by the class ratio.

\subsection{Analysis Methods}

In order to assess the transfer of ErrP categories within and between differing tasks, each sub-task acts as the source and target. We will refer to a source and target pair using the notation \trans{source}{target} (e.g., \trans{BGSInt}{OAOut}). Given this, there are 16 combinations where a given combination's source and target can share the same sub-task (e.g.\trans{BGSInt}{BGSInt}). We employ leave-one-group-out cross-validation (LOGOCV) to assess the transfer performance. LOGOCV is performed on each sub-task which has 10 trials each. During training, one trial is reserved as the testing data, while the remaining trials act as training data. When assessing the performance of the testing trial, performance is also measured on all other sub-task data. For example, BGSInt would produce 10 classifiers where each classifier would be assessed on the remaining sub-tasks (BGSObs, OAObs, OAOut) as well as the BGSInt test trial. While a variety of performance metrics are tracked during LOGOCV, we focus specifically on balanced accuracy (bACC), given as $\frac{1}{2}(\text{TPR}+\text{TNR})$ as it is a useful measure for imbalanced datasets \cite{iwane_inferring_2019, kim_handlingtransfer_2016}. We further break down bACC by analyzing its sub-components true positive rate (TPR) and true negative rate (TNR).

To assess the significance of the differences between scores, we utilize statistical significance testing. Testing for unequal variance between groups is done by running Levene and Bartlett tests in which both tests reject the null hypothesis. As such, we select Welch's T-test and the non-parametric version of the Tukey-Kramer multi-comparison test, the Games-Howell test. As we will see, $p$-values for the majority of the multi-comparison test are found to be significant. We suspect this is likely due to the large sample sizes \cite{durlak_how_2009}. For instance, LOGOCV produces the smallest sample size for baselines comparisons (\trans{source}{source}) which have a sample size of 90 (9 subjects$\times$10 trials$\times$1 test trial) while transfer comparisons (\trans{source}{target}) have a sample size of 900 (9 subjects$\times$10 trials$\times$10 tests trials from a different sub-task).

 Additionally, we utilize Hedges' $g^*$ effect size measure, which assumes unequal variance between groups and corresponding confidence intervals. The effect size (ES) will allow us to glean into the effect of a given result and further help to distinguish if a significant result might simply be finding small differences due to a large sample size \cite{durlak_how_2009}. Confidence intervals for the ES will provide further information regarding the variability and precision of results, regardless of $p$-values. Wider CIs indicate a result might be too variable to fully trust, and narrower CIs indicate the consistency of a result \cite{durlak_how_2009}. We follow Cohen's recommended rule of thumb for assessing the ``size" of an effect: very small ({\small$g^*<0.2$}), small ({\small$0.2\mkern1.5mu\leq\mkern1.5mu{g^*}\mkern1.5mu{<}\mkern1.5mu0.5$}), medium ({\small$0.5\mkern1.5mu\leq\mkern1.5mu{g^*}\mkern1.5mu{<}\mkern1.5mu0.8$}), and large ({\small$g^*\geq0.8$}).

%
\section{Results}\label{sec:results}

In the proceeding sections, we summarize our observations and discussions regarding the aforementioned research questions\footnote{Due to space constraints, not all plots and statistics are shown. To access more detailed plots, additional statistics, raw results, and source code, refer to \url{https://github.com/RL-BCI-Lab/deepbci/tree/cibci-2022}.}. Additionally, all $p$-values are assumed to be significant unless stated otherwise.

\begin{figure}[t]
  \centering
  \vspace{-0.5cm}
  \includegraphics[scale=.45]{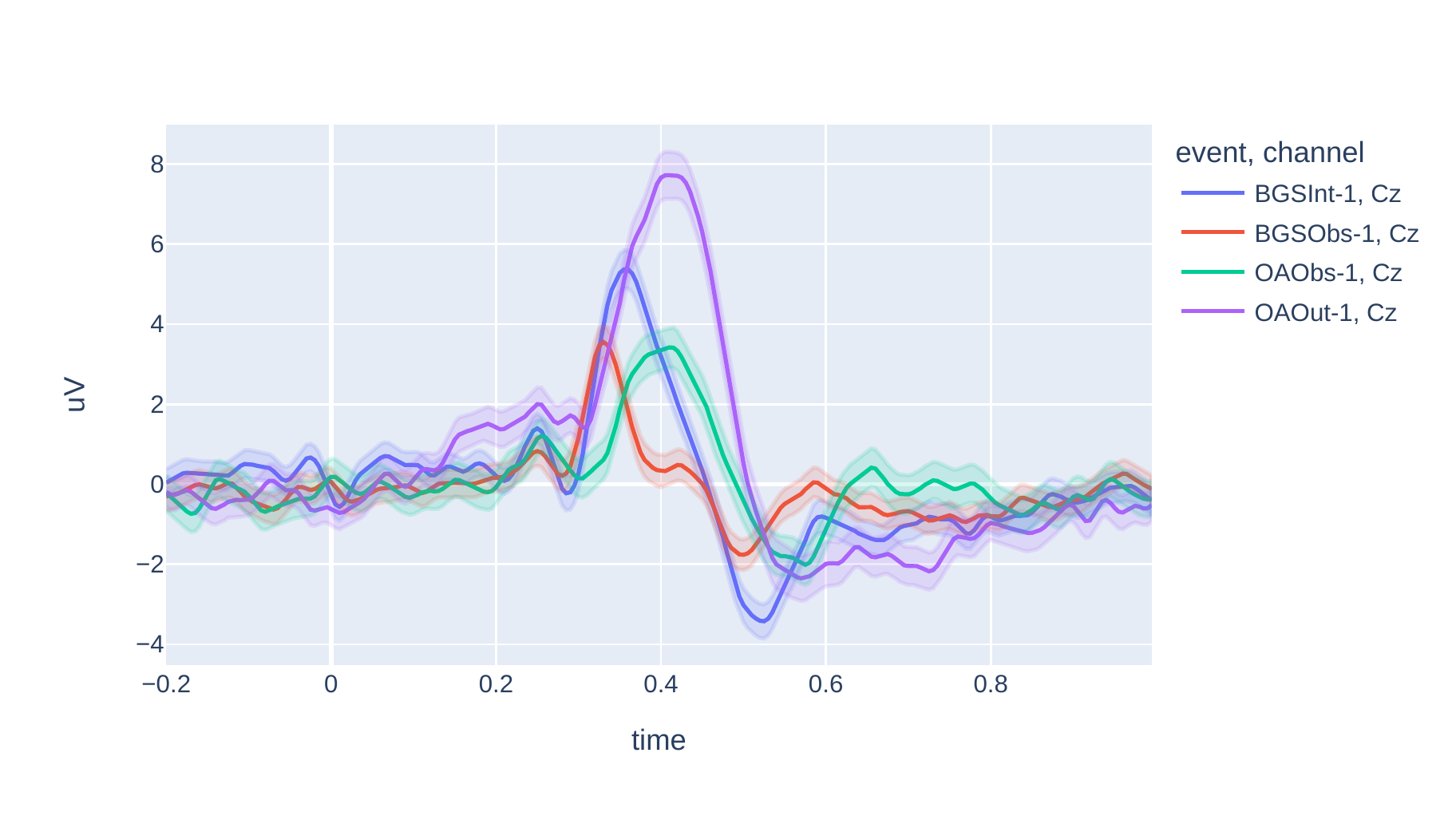}
  \vspace{-0.5cm}
  \caption{Grand averages for the Cz channel across all nine subjects for all four tasks with an IIR 0.1-30 Hz band-pass filter applied. The shaded regions represent the standard deviation across subjects.}
  \label{ga_pic}
  \vspace{-0.5cm}
\end{figure}

\paragraph{\textbf{Q1:} Do more ``distinct'' ErrP variations produce ``easier'' classification problems? }

\begin{figure*}[ht!]
    \centering
    \begin{subfigure}{\textwidth}
        \centering
        \includegraphics[width=0.25\textwidth]{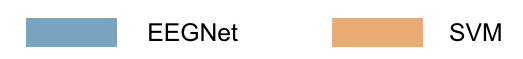}
    \end{subfigure}
    \begin{subfigure}{0.30\textwidth}
        \includegraphics[width=\linewidth]{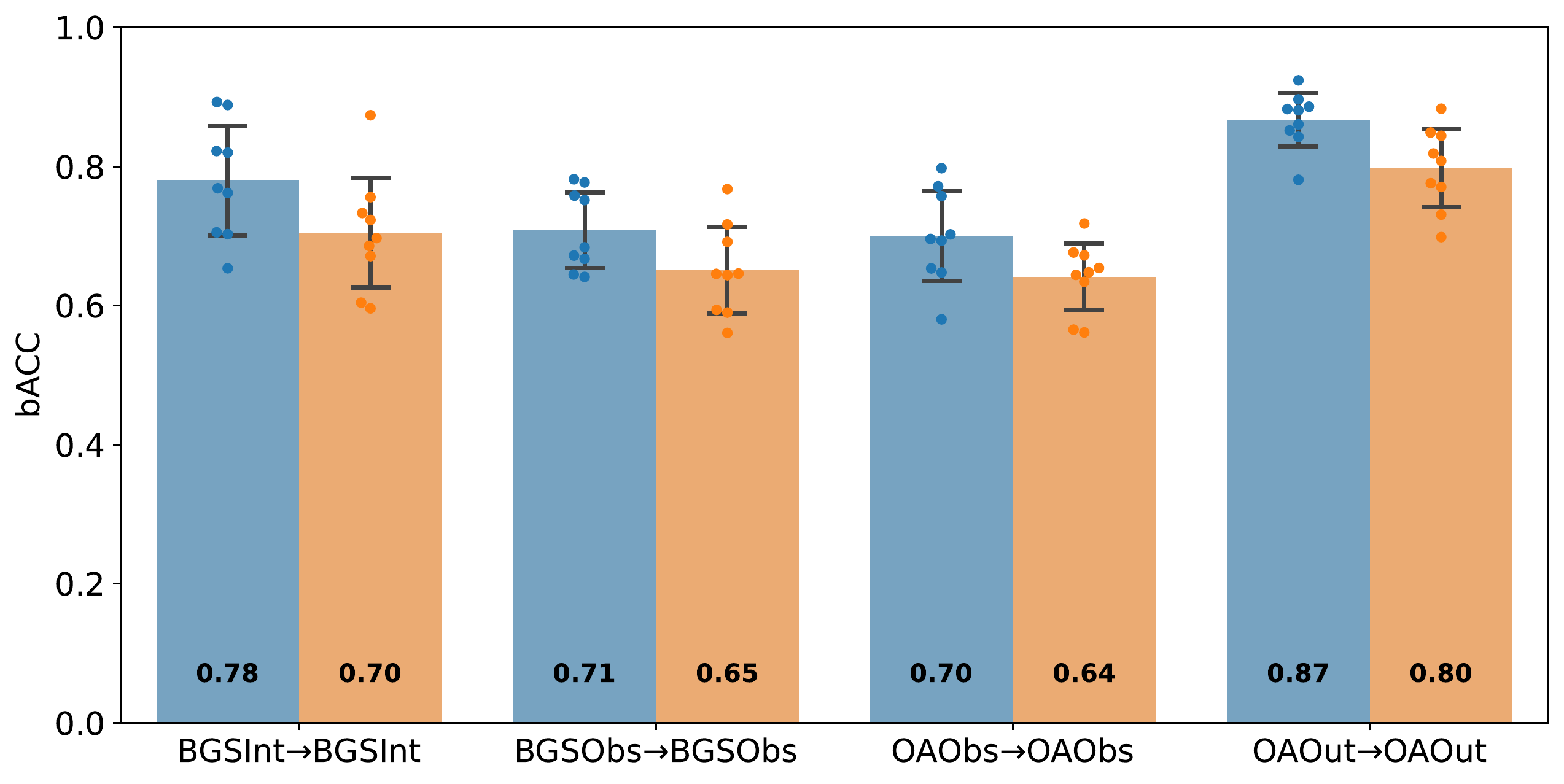}
        \caption{bACC} \label{fig:base_bacc}
    \end{subfigure}
    \begin{subfigure}{0.30\textwidth}
        \includegraphics[width=\linewidth]{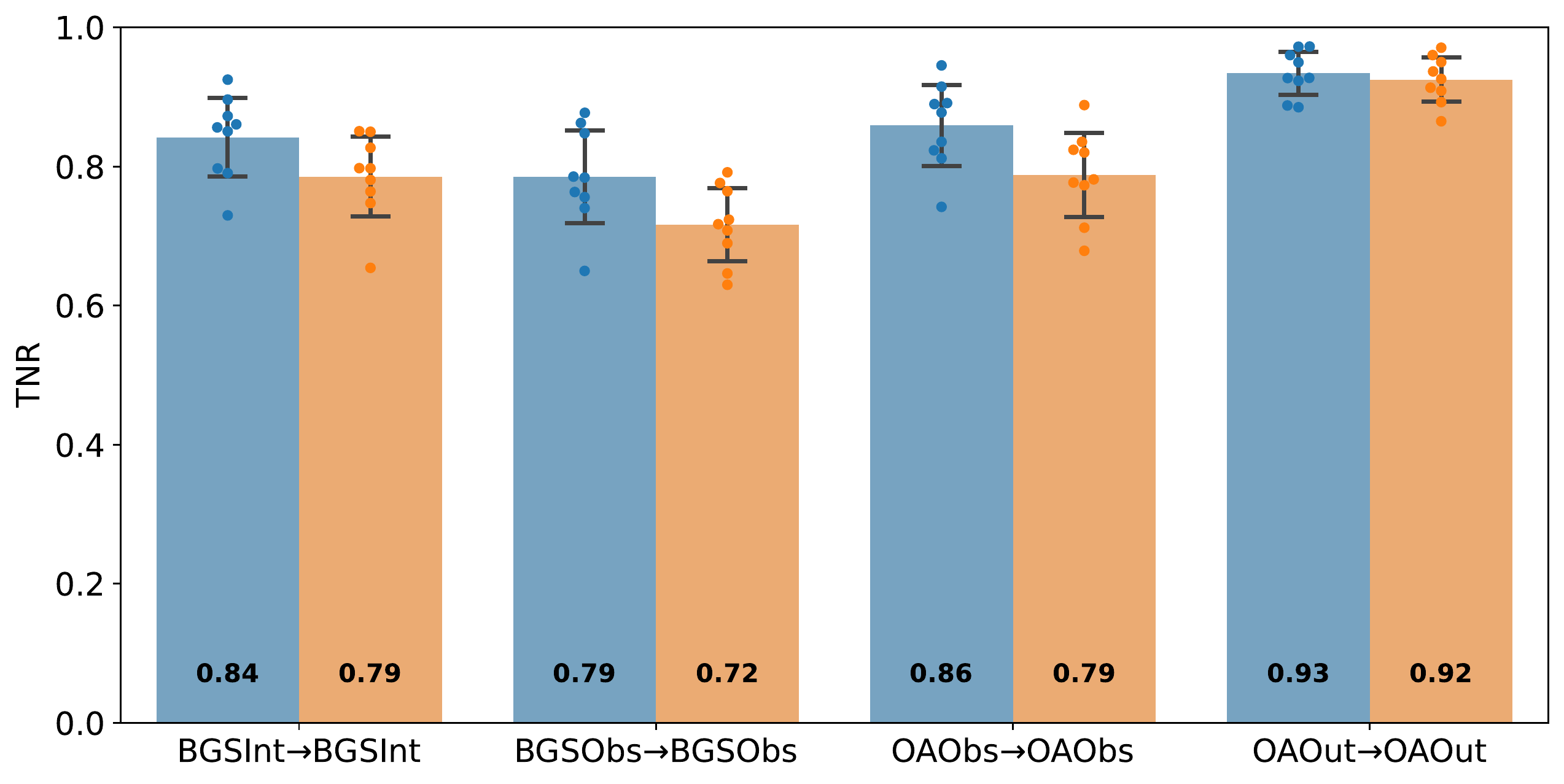}
        \caption{TNR} \label{fig:base_tnr}
    \end{subfigure}
    \begin{subfigure}{0.30\textwidth}
        \includegraphics[width=\linewidth]{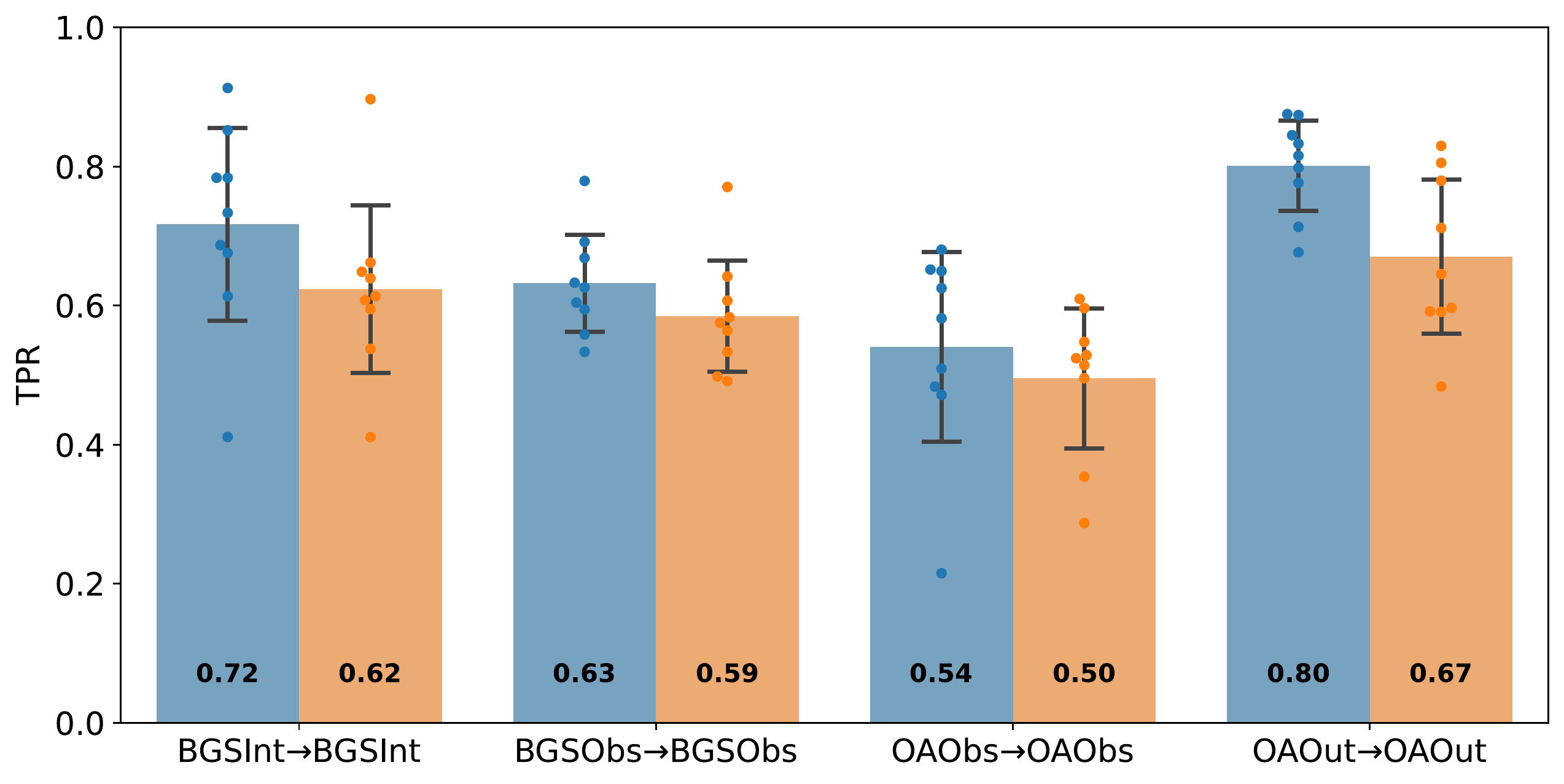}
        \caption{TPR} \label{fig:base_tpr}
    \end{subfigure}
    \vspace{-0.2cm}
    \caption{bACC, TNR, and TPR metrics for baseline comparisons.} 
    \label{fig:baseline_bACC}
    \vspace{-0.5cm}
\end{figure*}

To begin investigating this question, we first look at the grand signal averages. Recall, our initial hypothesis is that interactive tasks should produce more distinct signals due to potential increases in embodiment and awareness while observational tasks should produce less distinct signals due to decreases in the sense of embodiment and, potentially, awareness \cite{pavone_embodying_2016}. Figure~\ref{ga_pic} shows these grand signal averages (computed by averaging across all nine subjects) for each sub-task using the Cz channel. Although peak-to-peak artifact rejection was applied before averaging to remove any blatantly noisy epochs \cite{luck_erpbook_2014}, the data is still inherently quite noisy due to the BCI equipment and subject specific biological noise. Even so, the averages for the interactive tasks (i.e., BGSInt and OAOut) show more distinct Pe peaks (at approximately 400 ms) than their observational counterparts (i.e., BGSObs and OAObs). We hypothesize this increase is associated with interactive tasks inherently being more engaging than their observational counterparts, thus evoking higher degrees of attention and awareness in subjects \cite{orr_role_2011, iwane_invariability_2021}. However, among all four tasks, the ERN amplitude (at approximately 250 ms) remains relatively the same, with OAOut even having a positive amplitude. That being said, visually, there is a minor difference between BGSObs and BGSInt negative peaks. We suspect OAOut has a positive negative peak due to the noisier nature of this sub-task. For instance, additional cognitive and biological functions are likely required to perform the OA tasks compared to that of the BGS tasks. Additionally, the issue is likely further exaggerated by the inherently noisy ground truth of the OA task. This means subjects can often see collisions before one occurs, causing the ErrP to be elicited earlier than the actual collision (i.e., where the ErrP timestamp is thought to occur) \cite{spuler_continuouserrp_2015}. 

Due to the noisy nature of the data, the grand averages are not enough to produce conclusive results. Thus, we refer to the baseline comparisons (\trans{source}{source}) for indirect insights into which ErrP category is producing an easier classification problem and, in turn, could indicate more distinct signals. Using the Games-Howell test with the LOGOCV classification results, we first look at the bACC scores (\figref{fig:base_bacc}). When comparing the BGS tasks, BGSInt performs the best for both EEGNet (\summary{0.78}{0.71}{0.77}{0.47}{1.08}) with a medium to large ES and SVM (\summary{0.70}{0.65}{0.56}{0.26}{0.86}) with a small to large ES. Similarly, when comparing the OA tasks, OAOut performs best for EEGNet (\summary{0.70}{0.87}{-1.82}{-2.16}{-1.47})\footnote{Negative ESs indicate the direction of the effect. If the effect is negative, the second mean in a comparison is larger than the first.} and SVM (\summary{.64}{.80}{-1.66}{-2.0}{-1.32}) with large ESs. Similar results to bACC can be observed for TNR (\figref{fig:base_tnr}). However, for TPR (\figref{fig:base_tpr}), comparing the BGS tasks shows that BGSInt performs significantly better only for EEGNet (\summary{.72}{.63}{.48}{.18}{.77}) with a small to large ES. Meanwhile, SVM (\nonsummary{0.12}{0.62}{0.59}{0.23}{-0.06}{0.52}) shows non-significant results with corresponding very small to medium ES. More notably, when comparing OA tasks, OAOut performs significantly better for both EEGNet (\summary{0.54}{0.80}{-1.35}{-1.67}{-1.02}) with a large ES and SVM (\summary{0.50}{0.67}{-0.90}{-1.21}{-0.60}) with medium to large ES.

As we can see, the bACC baseline scores provide initial evidence that interactive data creates an easier classification problem across all scores for both classifiers. While these scores do not provide direct evidence for why the classification problem is easier, we further hypothesize that this is influenced by the distinct signals, in particular the Pe peaks, produced by interactive tasks. For instance, the fact that OAOut performs best across all metrics, specifically TPR, might indicate that OAOut generates the most distinct signals, which coincides with grand signal averages. In addition to the fact that TPR results for BGSInt are less conclusive, due to the subject-variability indicated by the larger standard deviation. This could mean that BGSInt, while having more distinct signals than its observational counterpart, reduces subject awareness/engagement compared to the OAOut sub-task. 

\paragraph{\textbf{Q2:} What role does the category of ErrP variation or task similarity play in transfer performance?}

Tangential to Q1, we now look at the effects of within task transfer (i.e., like task, differing ErrP categories) and within ErrP category transfer (i.e., differing task, like ErrP categories). In particular, we are looking to see how the easier classification problems (e.g., BGSInt and OAOut) perform when acting as either the source or target task.

\begin{figure*}[ht!]
    \begin{subfigure}{\textwidth}
        \centering
        \includegraphics[width=0.25\textwidth]{images/algo_legend.png}
    \end{subfigure}
    
     \begin{subfigure}{0.24\textwidth}
        \centering
        \includegraphics[width=1\linewidth]{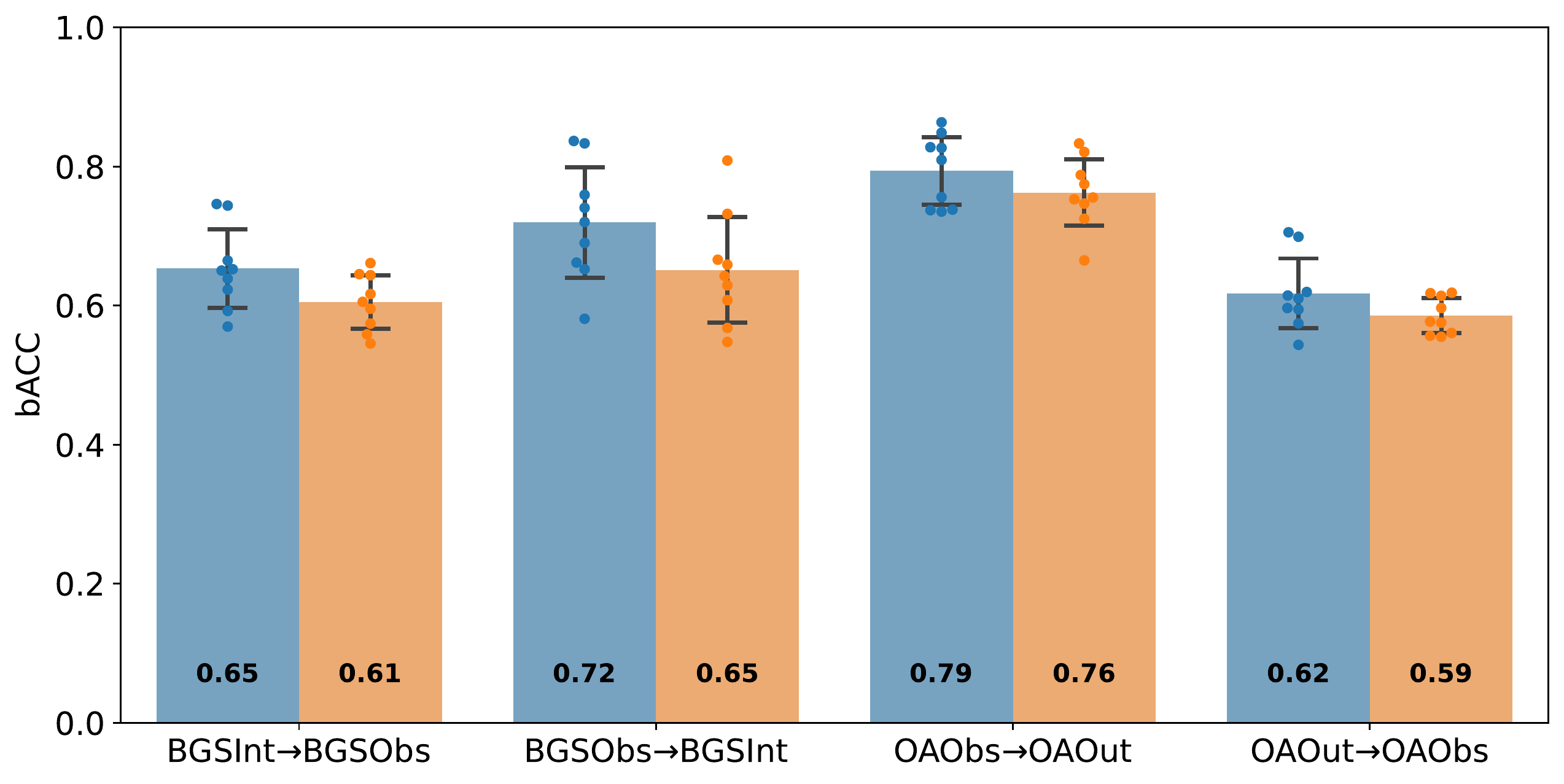}
        \caption{Within task (bACC)} \label{fig:within_task_bACC}
    \end{subfigure}
    \begin{subfigure}{0.24\textwidth}
        \centering
        \includegraphics[width=\columnwidth]{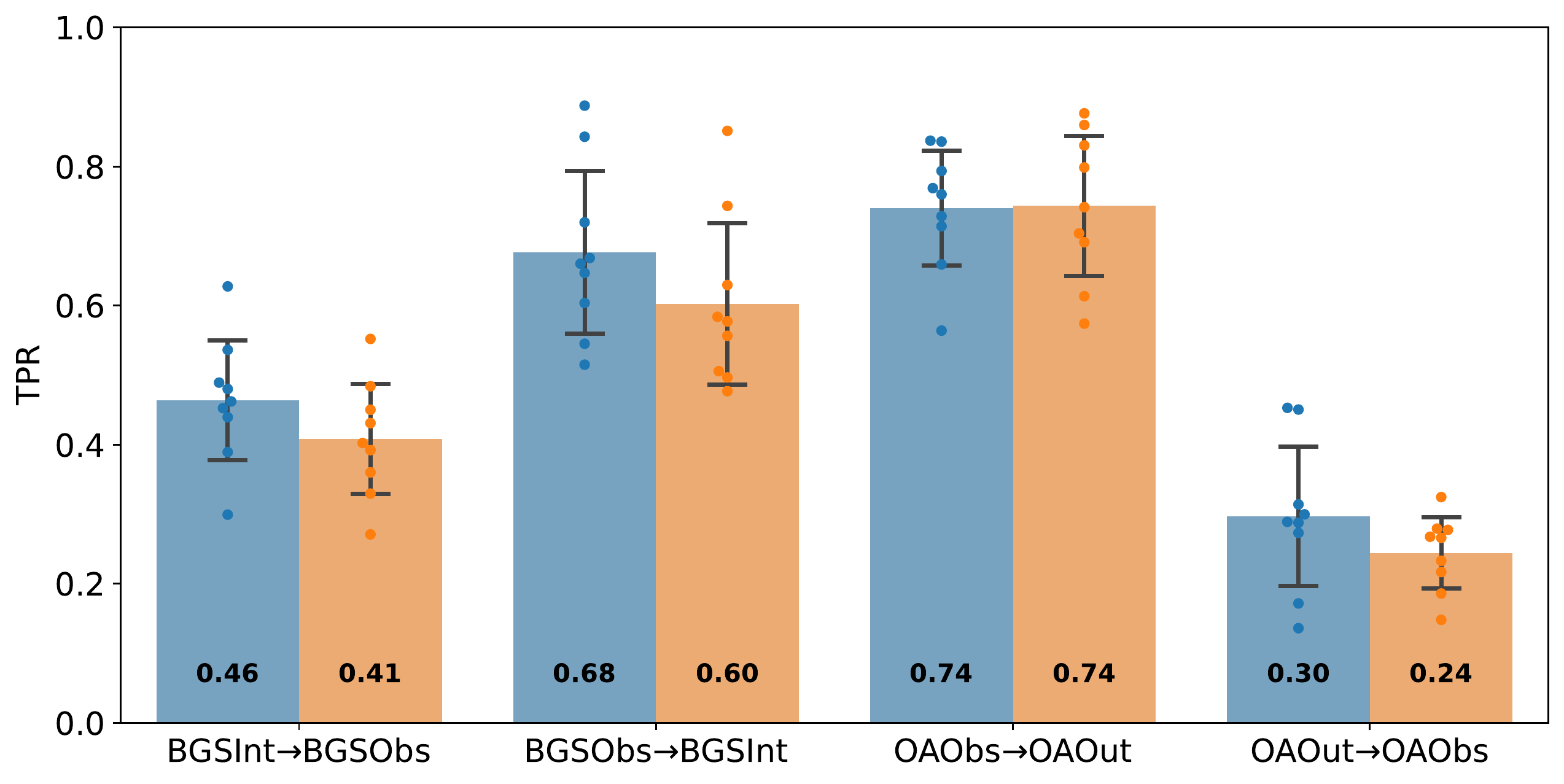}
        \caption{Within task (TPR)} \label{fig:within_task_TPR}
    \end{subfigure}
    \begin{subfigure}{0.24\textwidth}
        \centering
        \includegraphics[width=\linewidth]{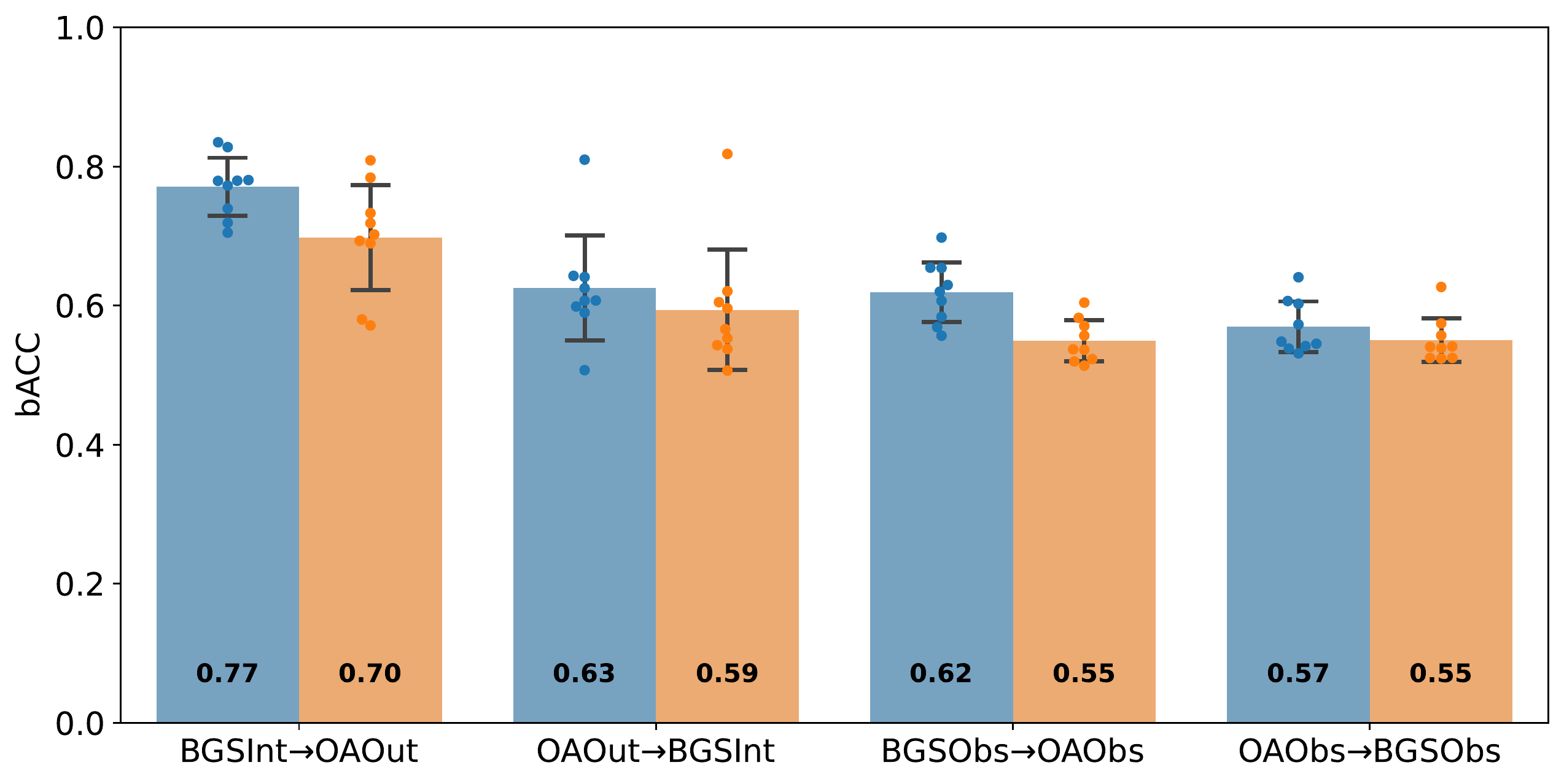}
        \caption{Within ErrP category (bACC)} \label{fig:with_category_bACC}
    \end{subfigure}
    \begin{subfigure}{0.24\textwidth}
        \centering
        \includegraphics[width=\linewidth]{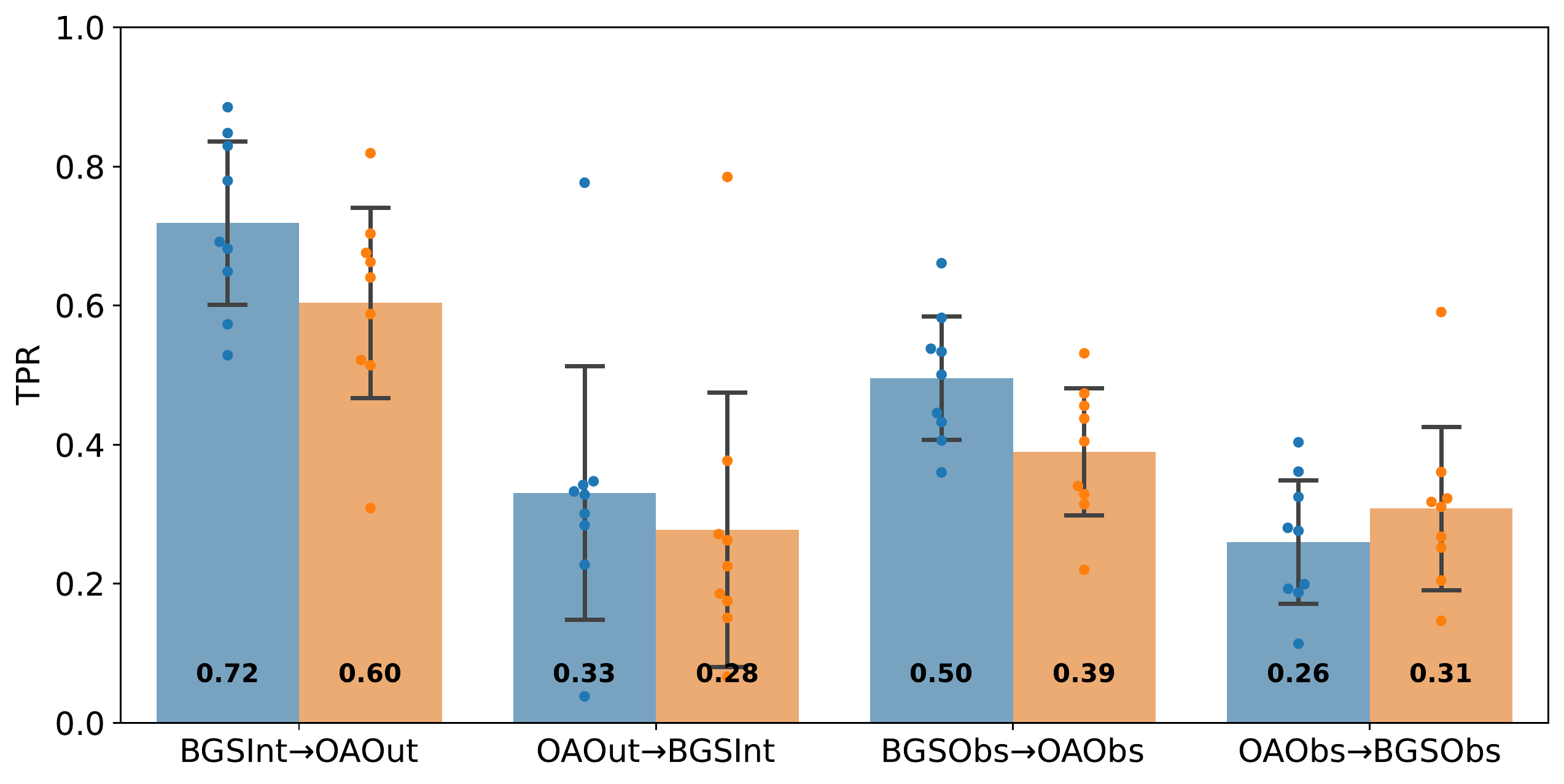}
        \caption{Within ErrP category (TPR)} \label{fig:with_category_TPR}
    \end{subfigure}
    \vspace{-0.2cm}
    \caption{Transfer performance for within task and ErrP category.} \label{fig:same_task_subtask}
    
    \begin{subfigure}{0.24\textwidth}
        \includegraphics[width=\linewidth]{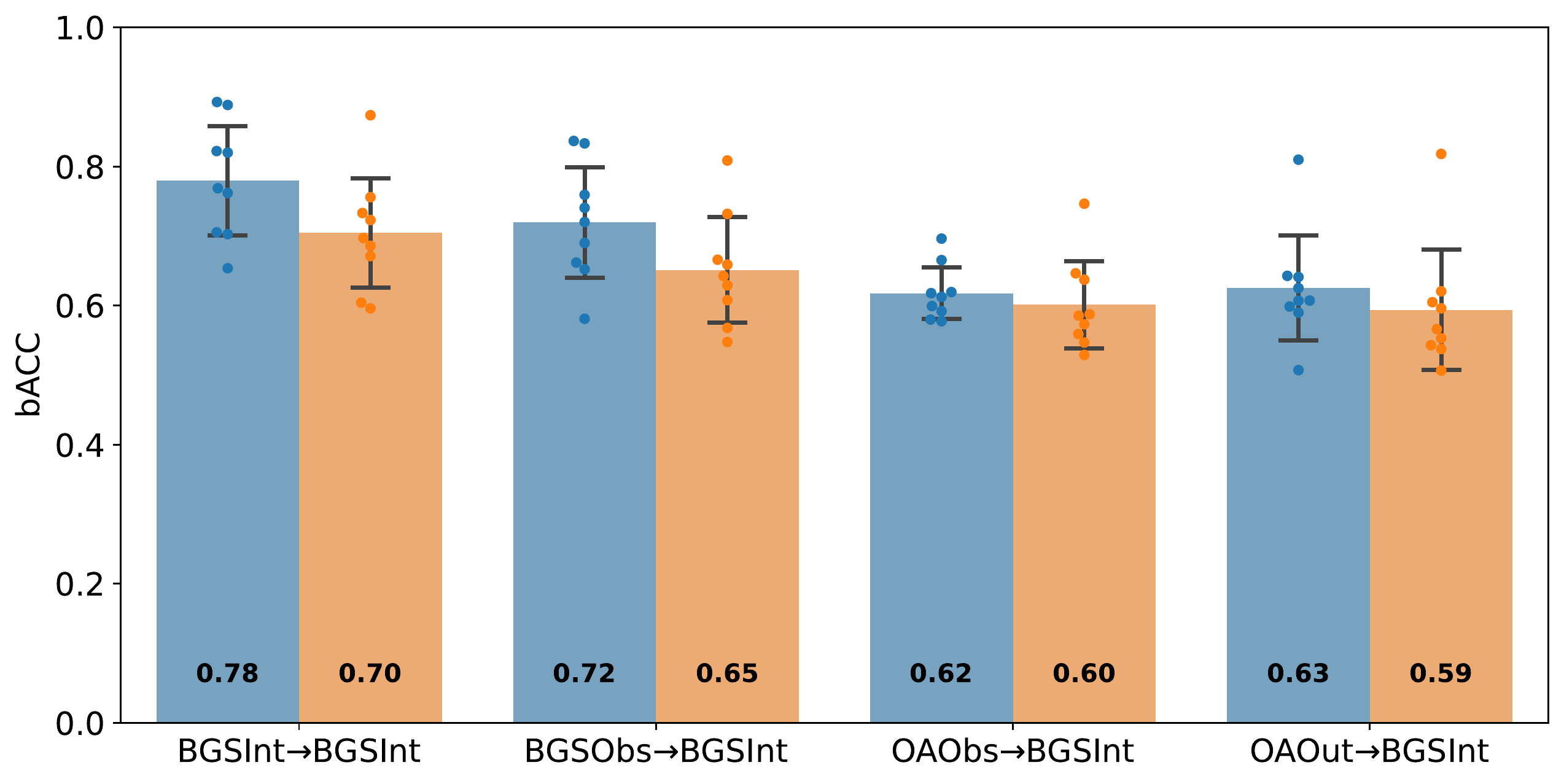}
        \caption{Target: BGSInt} \label{fig:test_bgsint}
    \end{subfigure}
    \begin{subfigure}{0.24\textwidth}
        \includegraphics[width=\linewidth]{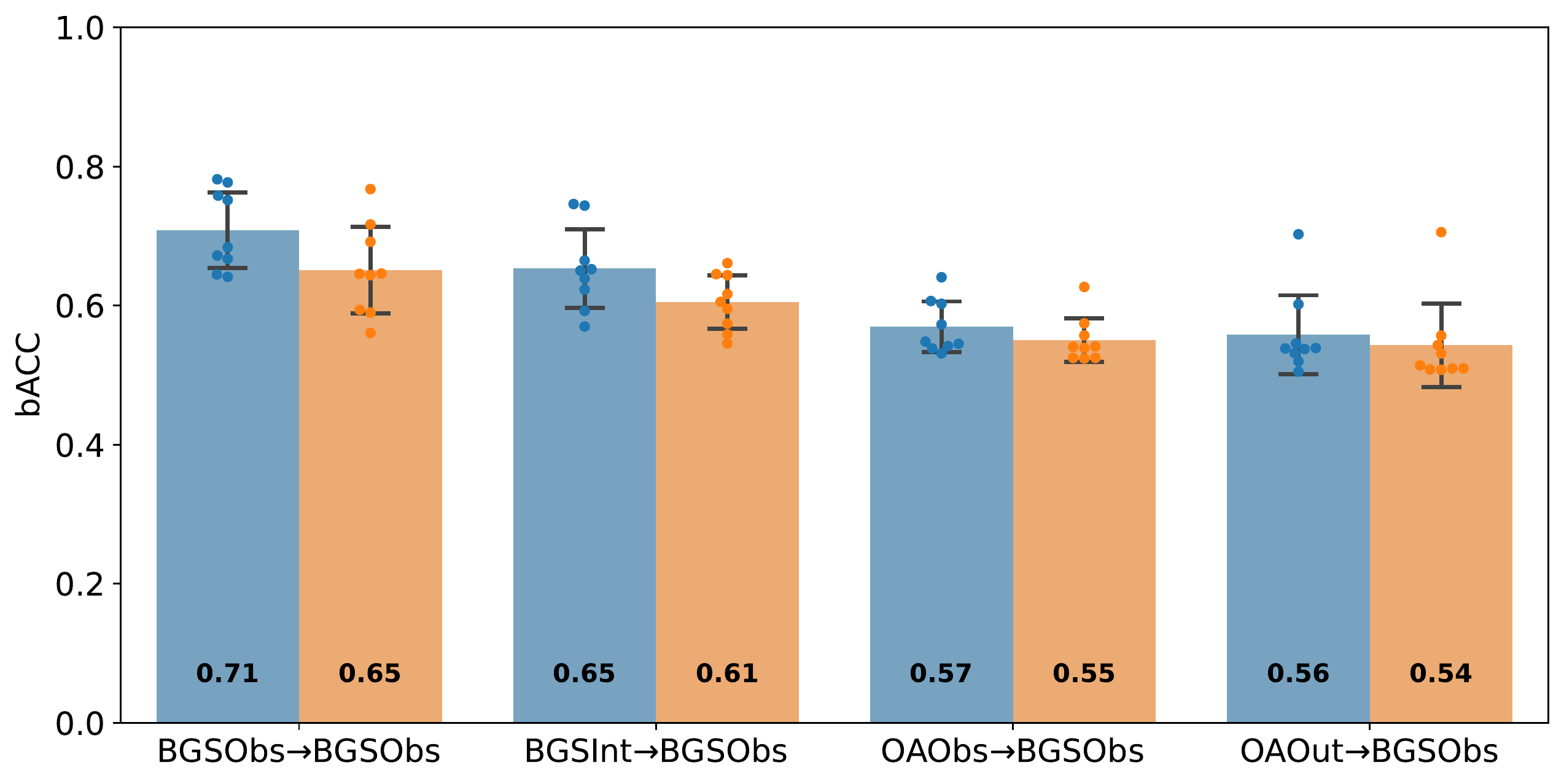}
        \caption{Target: BGSObs} \label{fig:test_bgsobs}
    \end{subfigure}
    \begin{subfigure}{0.24\textwidth}
        \includegraphics[width=\linewidth]{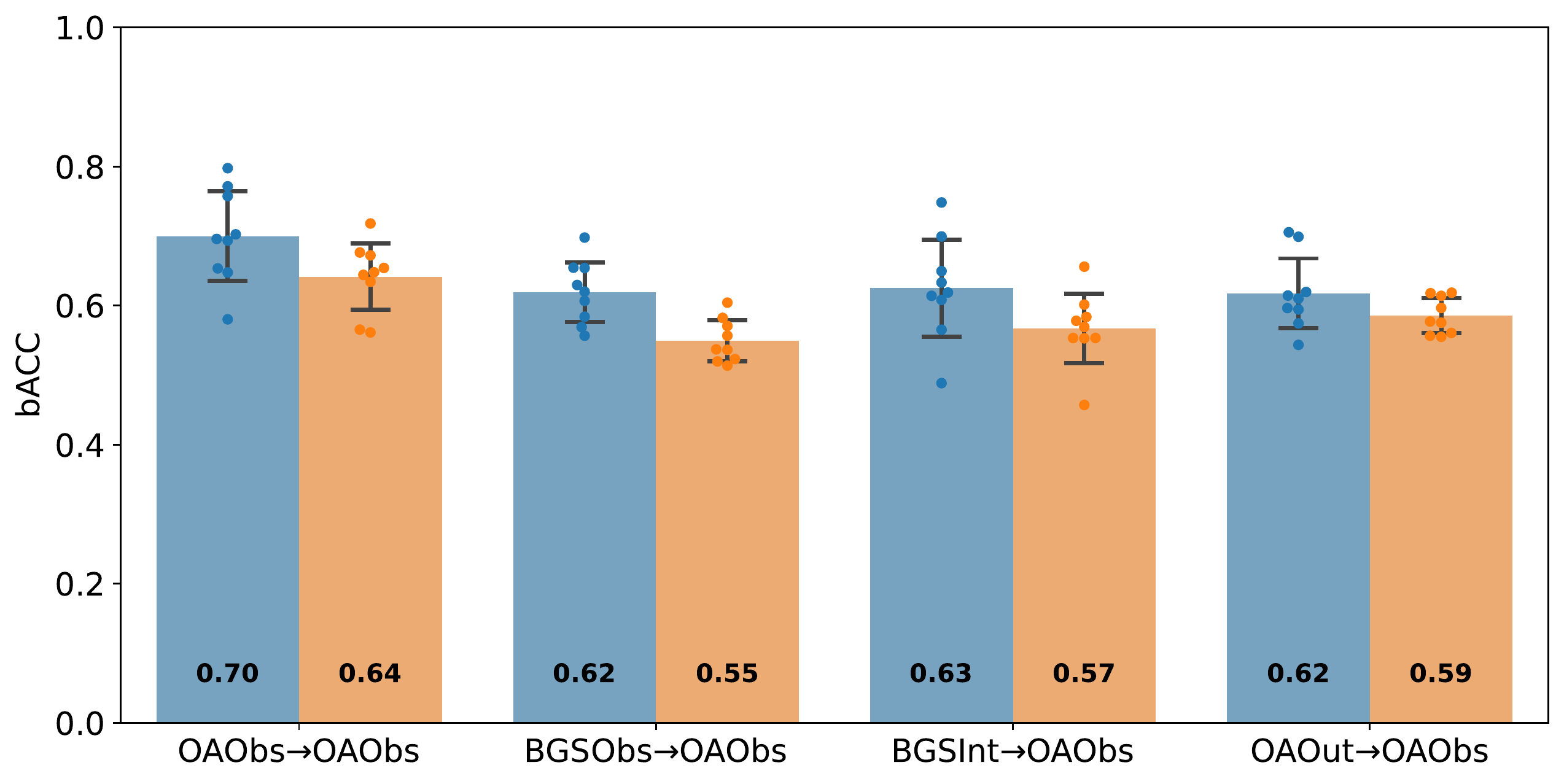}
        \caption{Target: OAObs} \label{fig:test_oaobs}
    \end{subfigure}
    \begin{subfigure}{0.24\textwidth}
        \includegraphics[width=\linewidth]{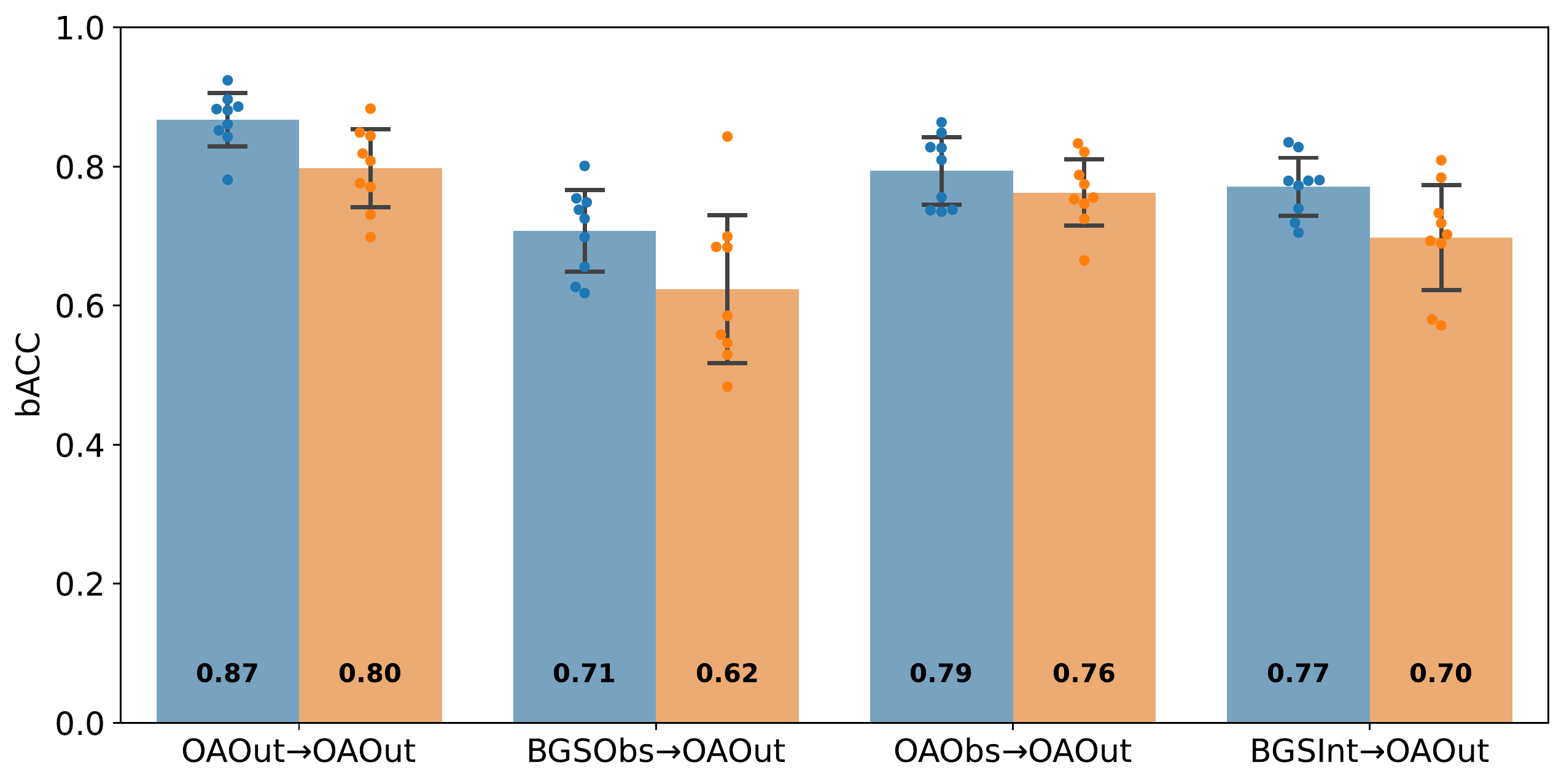}
        \caption{Target: OAOut} \label{fig:test_oaout}
    \end{subfigure}
    \vspace{-0.2cm}
    \caption{bACC transfer performance for same target tasks.} \label{fig:same_test_bACC}

    \begin{subfigure}{0.24\textwidth}
        \includegraphics[width=\linewidth]{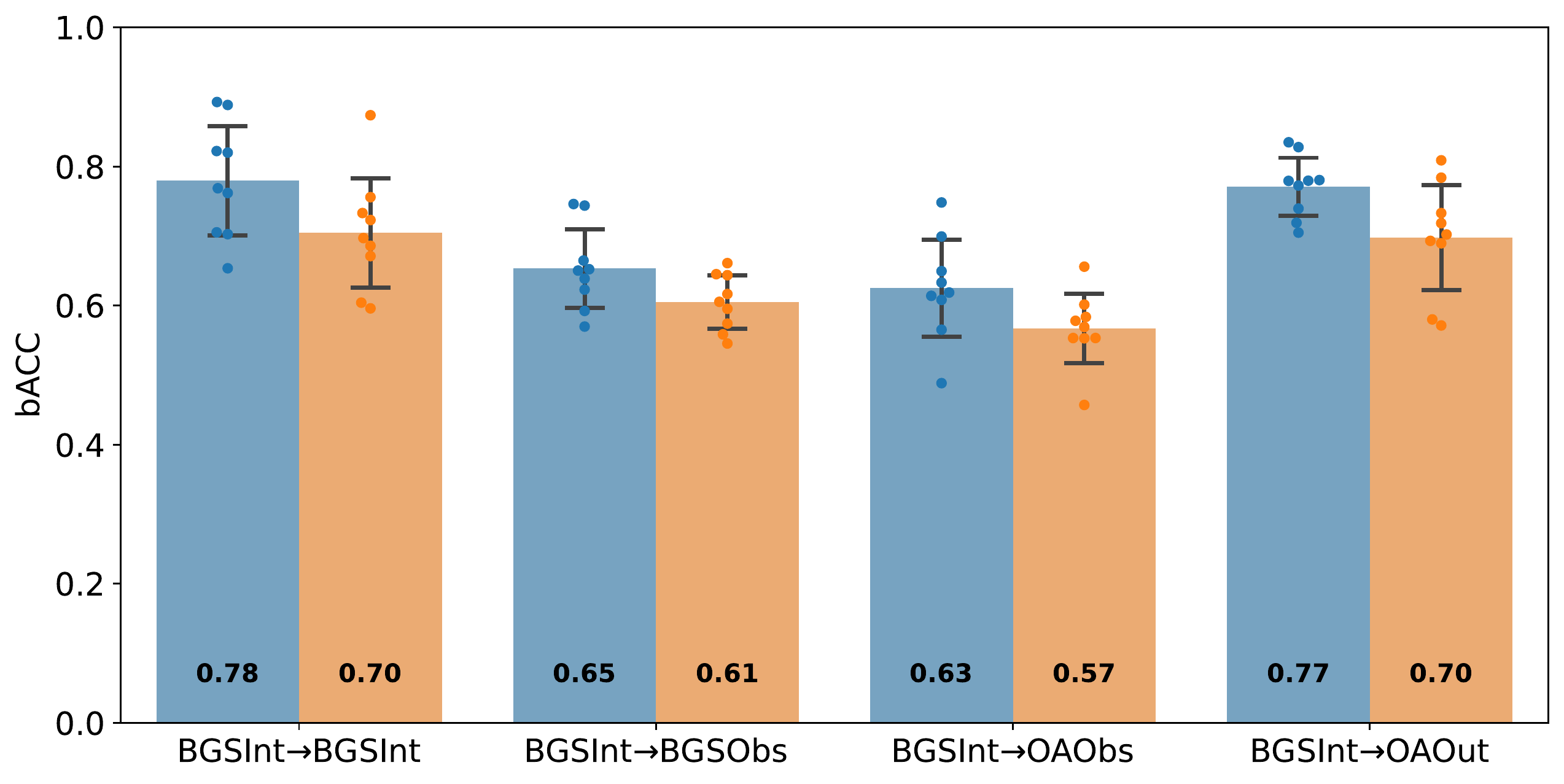}
        \caption{Source: BGSInt} \label{fig:src_bgsint}
    \end{subfigure}
    \begin{subfigure}{0.24\textwidth}
        \includegraphics[width=\linewidth]{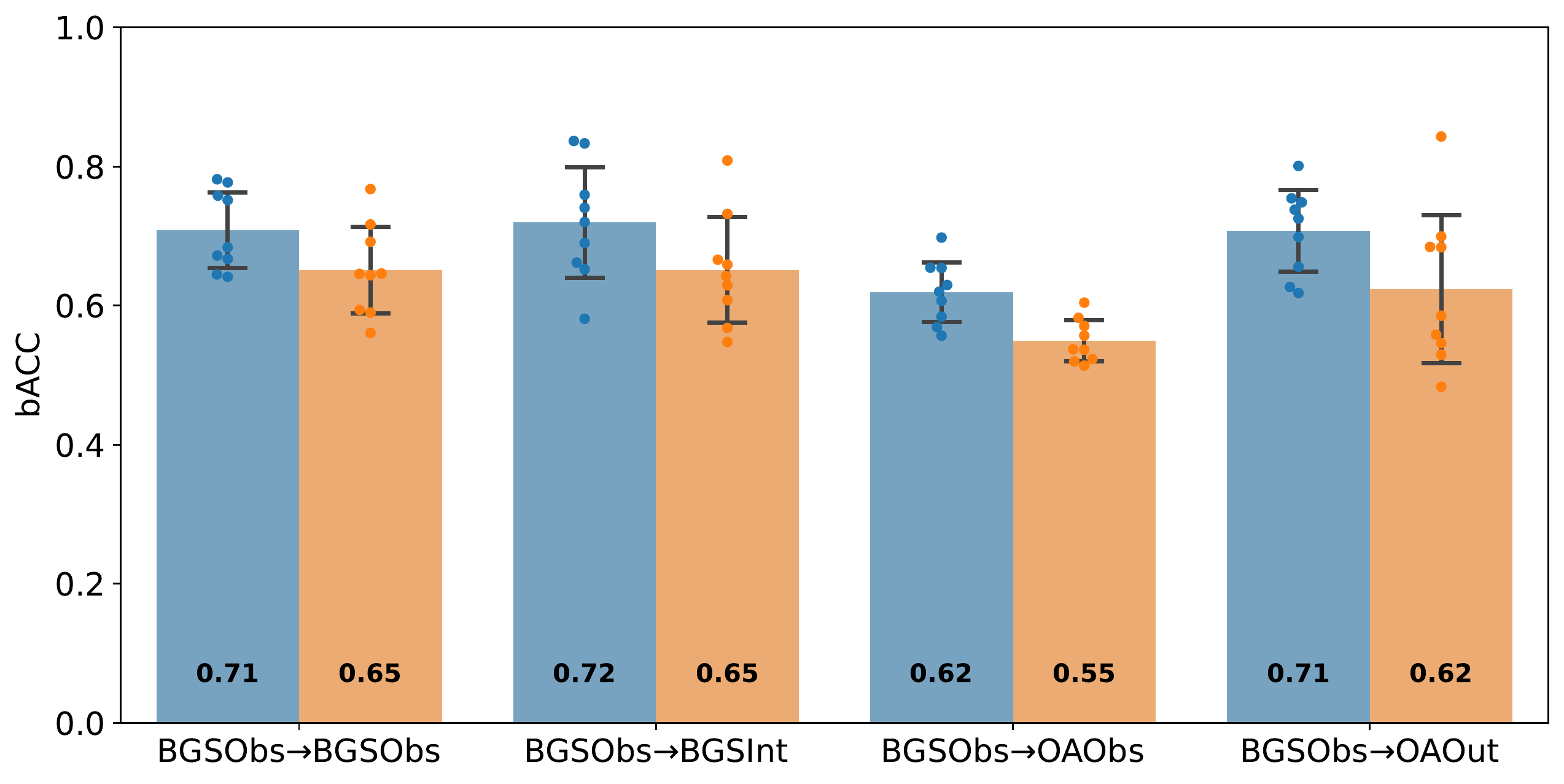}
        \caption{Source: BGSObs} \label{fig:src_bgsobs}
    \end{subfigure}
    \begin{subfigure}{0.24\textwidth}
        \includegraphics[width=\linewidth]{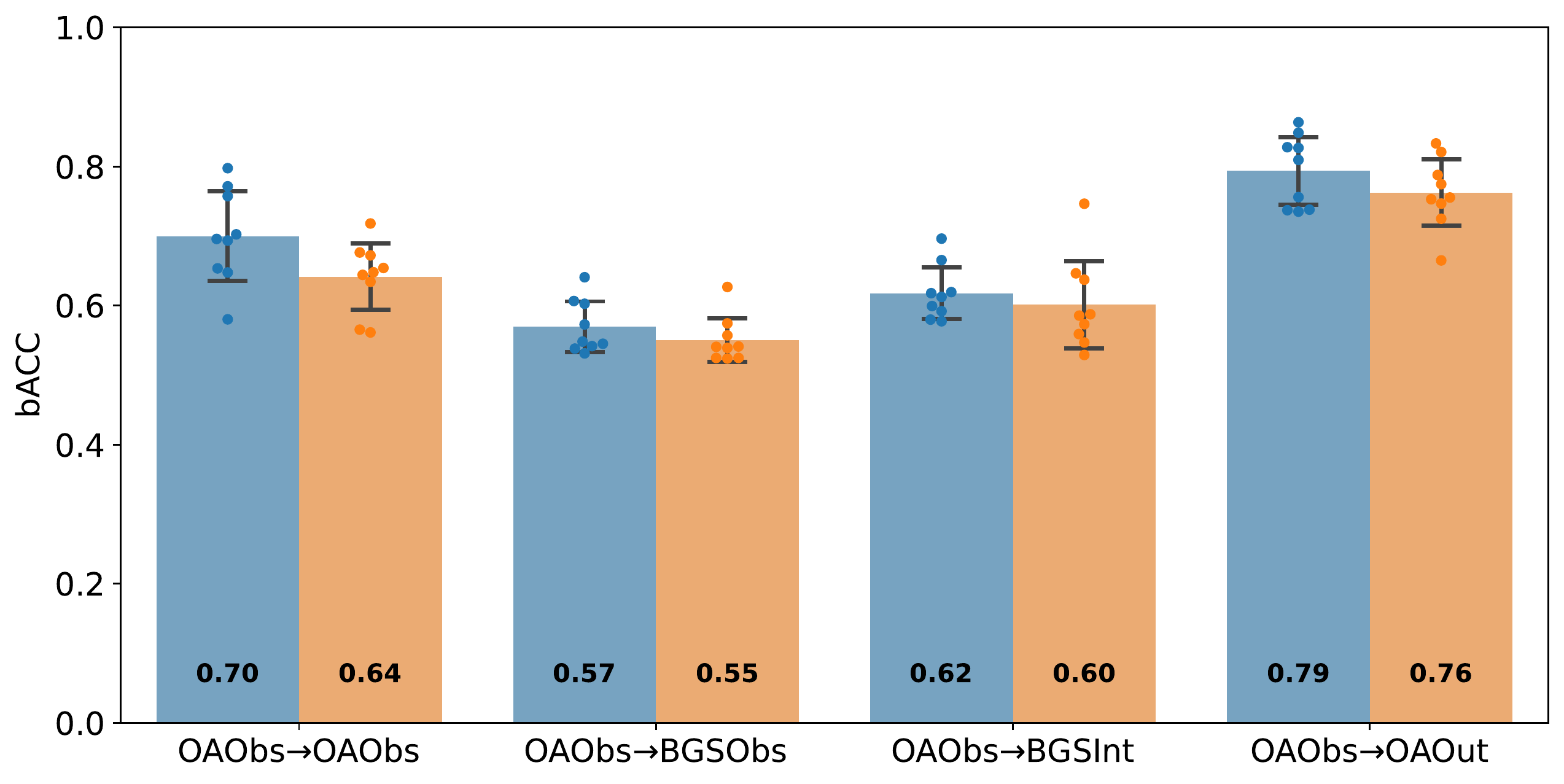}
        \caption{Source: OAObs} \label{fig:src_oaobs}
    \end{subfigure}
    \begin{subfigure}{0.24\textwidth}
        \includegraphics[width=\linewidth]{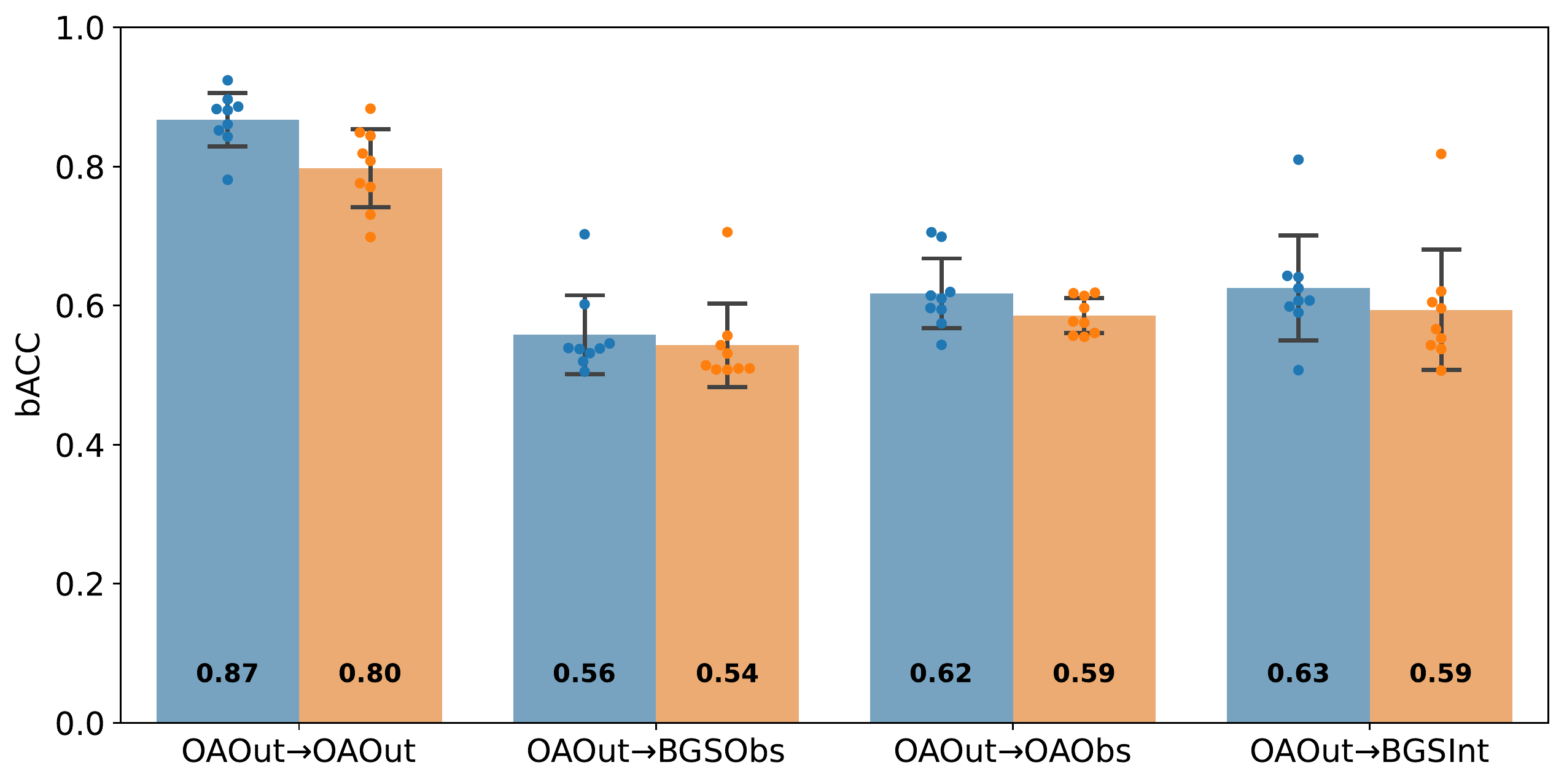}
        \caption{Source: OAOut} \label{fig:src_oaout}
    \end{subfigure}
    \vspace{-0.2cm}
    \caption{bACC transfer performance for same source tasks.} \label{fig:same_train_bACC}
    \vspace{-0.5cm}
\end{figure*}

First, we explore the within task comparison (\figref{fig:within_task_bACC} and \ref{fig:within_task_TPR}) by analyzing the LOGOCV results using Welch's T-test. Results point towards the interactive data producing the best transfer results when acting as the target, but performance suffers greatly when it acts as the source. This can be observed when comparing transfer within BGS as \trans{BGSObs}{BGSInt} outperforms \trans{BGSInt}{BGSObs} for both EEGNet (\summary{0.65}{0.71}{-0.68}{-0.78}{-0.59}) with a medium to large ES and SVM (\summary{0.61}{0.65}{-0.54}{-0.63}{-0.44}) with a small to medium ES. Likewise, when comparing transfer within OA, \trans{OAObs}{OAOut} outperforms \trans{OAOut}{OAObs} for EEGNet (\summary{0.79}{0.62}{1.87}{1.76}{1.98}) and SVM (\summary{0.76}{0.59}{2.23}{2.11}{2.35}) with both showing large ESs. TNR produces the opposite results as observed for bACC. For within BGS transfer, \trans{BGSInt}{BGSObs} outperforms \trans{BGSObs}{BGSInt} for EEGNet (\summary{0.84}{0.76}{0.9}{0.81}{1.0}) and SVM (\summary{0.80}{0.70}{1.67}{1.56}{1.78}) with both showing large ESs. For within OA transfer, \trans{OAOut}{OAObs} outperforms \trans{OAObs}{OAOut} for EEGNet (\summary{0.85}{0.94}{-1.43}{-1.54}{-1.32}) and SVM (\summary{0.78}{0.93}{-2.71}{-2.84}{-2.57}) where both show large ESs. The dip in bACC scores for interactive tasks arises from the TPR results. When comparing with BGS transfer, \trans{BGSObs}{BGSInt} outperforms \trans{BGSInt}{BGSObs} for EEGNet (\summary{0.46}{0.68}{-1.19}{-1.29}{-1.09}) and SVM (\summary{0.41}{0.60}{-1.23}{-1.33}{-1.12}) with large ESs. Lastly, when comparing with OA transfer, \trans{OAObs}{OAOut} outperforms \trans{OAOut}{OAObs} for EEGNet (\summary{0.74}{0.30}{2.29}{2.17}{2.41}) and SVM (\summary{0.74}{0.24}{3.04}{2.90}{3.18}) where both show large ESs.

Next, we explore the within ErrP category (i.e., between task) transfer comparison where we control for the ErrP category (\figref{fig:with_category_bACC} and \ref{fig:with_category_TPR}). Here, the problem starts to narrow as we see that OAOut in particular acts as the best target set. Additionally, we see the OA tasks have relatively low ErrP detection transfer (i.e., low TPR) but higher NoErrP detection (i.e., high TNR). Focusing on bACC results first. The interactive transfer, \trans{BGSInt}{OAOut} outperforms \trans{OAOut}{BGSInt} for EEGNet (\summary{0.77}{0.63}{1.52}{1.42}{1.63}) with a large ES and SVM (\summary{0.70}{0.59}{0.95}{0.86}{1.05}) with a medium to large ES. Observational transfer shows \trans{BGSObs}{OAObs} outperforms \trans{OAObs}{BGSObs} for EEGNet (\summary{0.62}{0.57}{0.64}{0.55}{0.74}) with a medium ES while the SVM shows non-significant results (\nonsummary{0.78}{0.55}{0.55}{-0.01}{-0.11}{0.08}). More interesting information can be gleaned from the TNR and TPR results. The TNR results show the opposite of the bACC results where the OA tasks increase TNR when acting as the source. For the interactive transfer,\trans{OAOut}{BGSInt} outperforms \trans{BGSInt}{OAOut} for EEGNet (\summary{0.82}{0.92}{-1.38}{-1.49}{-1.28}) and SVM (\summary{0.79}{0.91}{-1.84}{-1.95}{-1.72}) where both show large ESs. For the observational transfer,  \trans{OAObs}{BGSObs} outperforms \trans{BGSObs}{OAObs} for EEGNet (\summary{0.74}{0.88}{-1.70}{-1.81}{-1.60}) and SVM (\summary{0.71}{0.79}{-1.21}{-1.31}{-1.10}) where both show large ESs. On the other hand, TPR shows results similar to bACC.  For the interactive transfer, \trans{BGSInt}{OAOut} outperforms \trans{OAOut}{BGSInt} for EEGNet (\summary{0.72}{0.33}{1.77}{1.66}{1.88}) and SVM (\summary{0.60}{0.28}{1.44}{1.34}{1.54}) where both show large ESs. For the observational transfer, \trans{BGSObs}{OAObs} outperforms \trans{OAObs}{BGSObs} for EEGNet (\summary{0.50}{0.26}{1.35}{1.25}{1.46}) with a large ES and SVM (\summary{0.39}{0.31}{0.48}{0.39}{0.58}) with a small to medium ES. Given the results so far, a more specified question arises. Does observational data serve as good source data, as suggested in \cite{kim_handlingtransfer_2016}, or is interactive simply better target data? Furthermore, can we explore more into how the task confounds these interactions, in particular OAOut?

To explore this more specified question, we can look to comparisons where we control for the target (i.e., target remains constant across comparisons). This allows for all possible transfer combinations for a given target to be compared where baselines are also included. Using Games-Howell on the LOGOCV results, the bACC scores for same target performance can be seen in  \figref{fig:same_test_bACC}. We summarize the results by ranking their transfer performance. Table~\ref{tab:same_target_ranking} presents the rankings, which are determined by the ESs and $p$-values across the various combinations. Each same target grouping contains 12 comparisons (6 for each classifier). The $>$ sign indicates a group's performance is better than another (``better'' indicates significant results or medium to large ES) while $\sim$ indicates a group's performance is similar to that of another (``similar'' indicates non-significant results, very small to small ES, or conflicting results). Additionally, rankings of the baseline performance are given as supplementary information. 

For the bACC scores, rankings remain relatively intuitive. As might be expected, the baseline (\trans{source}{source}) always outperforms the transfer (\trans{source}{target}) classification. The rankings also suggest that the within task counterpart of the target, not the within ErrP category counterpart, often leads to the 2nd best classification performance (i.e., best transfer performance). The exception here being OAObs, which has a three-way tie for 2nd between all the other sub-tasks. For the BGS targets, the 3rd place is always a tie between the OA tasks. This indicates that having the same ErrP variation might not matter, but we suspect this might actually be due to poor OA performance as a source task in general. Finally, when OAOut is the target, we see the clearest ranking.

Again, the TPR and TNR show intriguing results. TNR shows very consistent rankings across all transfers. In particular, OAOut always produces the highest TNR results followed by a tie between BGSInt and OAObs, with BGSObs always producing the worst TNR scores. TPR tells a less clear story when the target is an interactive dataset. For instance, when using BGSInt as target data, ties arise for 2nd and 3rd place. Furthermore, when using OAOut as the target, the ranking can not be clearly established as both classifiers produce conflicting results between one another and the variation between subjects is quite large.  

\begin{table}[t]
    \scriptsize
    \caption{Transfer rankings when using the same target task with baseline performance rankings included for reference.}\label{tab:same_target_ranking}
    \vspace{-0.3cm}
    \center
    \begin{tabular}{|c|c|c|C{1cm}|}
    \hline
     Metric & Target & Source Task Rank & Baseline Rank \\
    \hline
    bACC & BGSInt & BGSInt $>$ BGSObs $>$ OAOut$\sim$OAObs & 2\\
    bACC & BGSObs & BGSObs $>$ BGSInt $>$ OAOut$\sim$OAObs & 3\\
    bACC & OAObs & OAObs $>$ OAOut$\sim$BGSInt$\sim$BGSObs & 3\\
    bACC & OAOut & OAOut $>$ OAObs $>$ BGSInt $>$ BGSObs & 1\\
    \hline
    TNR & BGSInt & OAOut $>$  BGSInt$\sim$OAObs $>$ BGSObs & 2 \\
    TNR & BGSObs & OAOut $>$  BGSInt$\sim$OAObs $>$ BGSObs & 3\\
    TNR & OAObs & OAOut $>$  BGSInt$\sim$OAObs $>$ BGSObs & 2\\
    TNR & OAOut & OAOut $>$  BGSInt$\sim$OAObs $>$ BGSObs & 1\\
    \hline
    TPR & BGSInt & BGSInt$\sim$BGSObs $>$ OAOut$\sim$OAObs & 2\\
    TPR & BGSObs & BGSObs $>$ BGSInt $>$ OAObs $>$ OAOut & 2\\
    TPR & OAObs &  OAObs $>$ BGSObs $>$ BGSInt $>$ OAOut & 3\\
    TPR & OAOut &  OAOut$\sim$OAObs$\sim$BGSInt$\sim$BGSObs & 1\\
    \hline
    \end{tabular}
    \vspace{-0.3cm}
\end{table}

\begin{table}[t]
    \scriptsize
    \caption{Transfer rankings when using the same source with baseline performance rankings included for reference.}\label{tab:same_source_ranking}
    \vspace{-0.3cm}
    \center
    \begin{tabular}{|c|c|c|C{1cm}|}
    \hline
     Metric & Source & Target Task Rank & Baseline Rank \\
    \hline
    bACC & BGSInt & BGSInt$\sim$OAOut $>$ BGSObs$\sim$OAObs & 2\\
    bACC & BGSObs & BGSObs$\sim$OAOut$\sim$BGSInt $>$ OAObs & 3\\
    bACC & OAObs & OAOut $>$ OAObs $>$ BGSInt $>$ BGSObs & 3\\
    bACC & OAOut & OAOut $>$ OAObs$\sim$BGSInt $>$ BGSObs & 1\\
    \hline
    TNR & BGSInt & BGSObs$\sim$BGSInt$\sim$OAObs$\sim$OAOut & 2\\
    TNR & BGSObs & BGSObs$\sim$OAOut$\sim$BGSInt$\sim$OAObs & 3\\
    TNR & OAObs & OAObs$\sim$BGSInt$\sim$OAObs$\sim$BGSObs & 2\\
    TNR & OAOut & OAObs$\sim$OAOut$\sim$BGSObs$\sim$BGSObs & 1\\
    \hline
    TPR & BGSInt & BGSInt$\sim$OAOut $>$ BGSObs$\sim$OAObs & 2\\
    TPR & BGSObs & BGSObs$\sim$OAOut$\sim$BGSInt $>$ OAObs & 2\\
    TPR & OAObs &  OAOut $>$ OAObs $>$ BGSInt $>$ BGSObs & 3\\
    TPR & OAOut &  OAOut $>$ OAObs$\sim$BGSInt $>$ BGSObs & 1\\
    \hline
    \end{tabular}
    \vspace{-0.5cm}
\end{table}

For additional exploration, the reverse comparisons can be made where we control for the source set  (i.e., the source remains constant across comparisons). This allows for all possible transfer combinations for a given source to be compared where baselines are also included. Using Games-Howell on the LOGOCV results, the bACC results for the same target performance can be seen in \figref{fig:same_train_bACC}. Table~\ref{tab:same_source_ranking} shows all the rankings which are determined the same as in Table~\ref{tab:same_target_ranking}.

The bACC results shows that OAOut acts as one of the best target sets that even rivals baseline performance. For example, \trans{BGSInt}{OAOut} and \trans{BGSObs}{OAOut} show no significant difference with the baselines (e.g., \trans{BGSInt}{BGSInt} or \trans{BGSObs}{BGSObs}). Furthermore, when using OAObs as the source, the performance of \trans{OAObs}{OAOut} is better than the baseline \trans{OAObs}{OAObs}. Interestingly, these same results are not fully reflected for BGSInt, suggesting something particularly unique about OAOut. It is also worth noting that the OAOut results (\figref{fig:src_oaout}) clearly show a drastic drop in transfer performances when compared to its baseline. Such a drop is not seen in other sub-tasks' transfer performance with only BGSInt being the next closest. Moreover, TNR results show that baseline TRN performance is approximately maintained across transfer such that transfer scores are very similar to the baseline (i.e., non-significant differences or very small to small ESs). Similarly, TPR results show that OAOut performs similarly to the baselines, if not better. TPR additionally shows the same ties that are reflected in the bACC scores.

\paragraph{\textbf{Q3:} Following Q1 and Q2, what practical implications can be extracted?}

First, baseline comparisons clearly show that easier classification problems correspond with interactive tasks as BGSInt and OAOut received top classification performance. In particular, OAOut seems to perform, all around, the best. Now, what exactly causes these interactive tasks to be ``easier'' remains undetermined. As we mentioned, we suspect the distinction of the signal influences how easy a task is. Yet, what defines a distinct signal remains relatively unproven. In particular, we hypothesized that awareness/engagement and embodiment are significant contributors as suggested by prior works \cite{iwane_invariability_2021, orr_role_2011}. Sadly, due to the noisy aspects of our data, it is hard to make direct connections to the effects of embodiment and awareness to the grand signal averages. Regardless, we suspect that the emphasised Pe peaks of BGSInt and OAout, whether caused by increased awareness or not, lead to this increase in classification ability. Yet, further investigation and replication of this is required and serves as a good direction for future work.

When it comes to transfer, it becomes clearer that these ``easier'' classification problems perform well when used as the target rather than the source. One could hypothesize that this is due to the distinct nature of the interactive data, causing classifiers to overfit and, in turn, generalize poorly to the less distinct signals, such as the observational data. In particular, BGSInt performs similarly to and often better than OAOut at transfer when acting as the source task. This can be clearly seen by comparing \figref{fig:src_bgsint} and \figref{fig:src_oaout}. In general, the OA tasks seem to clearly demonstrate this idea of classifiers learning patterns that cannot generalize properly to differing tasks or even the same task. Another clear example of this is how \trans{OAOut}{OAObs} performs significantly worse than \trans{OAObs}{OAOut} (\figref{fig:within_task_bACC}), in particular when looking at the TPR (i.e., ErrP detection). Furthermore, for BGS tasks, it seems that the within task transfer outperforms the within ErrP category transfer. However, this is slightly contradicted by the OA tasks. For example, when OAObs is the target (\figref{fig:test_oaobs}) all source tasks perform similarly. Additionally, when OAOut is the target, comparing \trans{OAObs}{OAOut} and \trans{BGSInt}{OAOut} shows a small ES for EEGNet and a medium ES for SVM. 

Yet, when choosing the source task, things become less straightforward. The results at least show that task similarity guarantees a certain level of performance drop in comparison to baseline scores (e.g., 5-8 points drop in bACC is seen in \figref{fig:same_test_bACC}). However, depending on the task and the distinctive nature of the ErrP signal, the performance drop can fluctuate. BGSInt seems to be a good middle ground source task as the transfer performance is similar to that of \trans{OAOut}{OAObs} and \trans{OAObs}{OAOut} (\figref{fig:same_test_bACC}). Additionally, BGSInt has the added benefit that it is a relatively controlled task that allows for an abundance of data to be collected. Using different variations of the BGSInt task, as seen in prior works \cite{abu_errpinvariance_2017, iturrate_errptask_2013, iwane_invariability_2021}, to increase complexity and make it more engaging (i.e., increasing awareness) might lead to an increase in baseline and transfer performance without suffering the degradation in transfer performance OAOut has.

\section{Conclusion}\label{sec:conclusion}

 This paper provides an initial exploration into the transferability of the interactive and observational ErrP categories between similar and differing tasks. We do so by empirically exploring the impact of within task and ErrP category transfer with a focus on the ability for easier classification problems to transfer. We find that interactive tasks produce easier classification problems. However, easier classification problems like OAOut risk providing poor source data. Even so, we do believe BGSInt serves as a good general source task for both BGS and OA tasks. As this line of work aims to better understand the role of data in ErrP transfer performance, this understanding can then be used to help determine what source-target task design is needed such that there is a minimal drop in classifier performance. Yet, there is much more work to be done to account for other potential confounds such as the role async classification, more refined exploration into contributors to ErrP variation ``distinctiveness'' and other possible contributors such as back-to-back errors \cite{ferrez_errp_2008}, and whether these results can be further replicated across other tasks and classifiers.
 
\bibliographystyle{ieeetr}
\bibliography{manuscript}

\end{document}